\documentclass[prb,amssymb,twocolumn]{revtex4}

\usepackage{graphicx}% Include figure files
\usepackage{txfonts}
\usepackage{epsfig}

\def\avg#1{\left\langle#1\right\rangle}
\def\bra#1{\left\langle#1\right|}
\def\ket#1{\left|#1\right\rangle}

\def\abs#1{\left|#1\right|}
\def\kc#1{\left(#1\right)}
\def\kd#1{\left[#1\right]}

\def\be{\begin{equation}}       \def\ee{\end{equation}}
\def\bea{\begin{eqnarray}}      \def\eea{\end{eqnarray}}
\def\ba{\begin{array} }
\def\ea{\end{array} }
\def\bnum{\begin{enumerate} }
\def\enum{\end{enumerate}}

\def\nn{\nonumber}

\def\=>{\Rightarrow}
\def\>{\rightarrow}

\def\eye{\mathbb{I}}

\def\vect#1{\kc{\ba{c}#1\ea}}

\begin{document}
\title{Exact holographic mapping and emergent space-time geometry}

\author{Xiao-Liang Qi}
\affiliation{ Department of Physics, Stanford University, Stanford, California 94305, USA}
\date{\today}
\begin{abstract}
In this paper, we propose an {\it exact holographic mapping} which is a unitary mapping from the Hilbert space of a lattice system in flat space (boundary) to that of another lattice system in one higher dimension (bulk). By defining the distance in the bulk system from two-point correlation functions, we obtain an emergent bulk space-time geometry that is determined by the boundary state and the mapping. As a specific example, we study the exact holographic mapping for $(1+1)$-dimensional lattice Dirac fermions and explore the emergent bulk geometry corresponding to different boundary states including massless and massive states at zero temperature, and the massless system at finite temperature. We also study two entangled one-dimensional chains and show that the corresponding bulk geometry consists of two asymptotic regions connected by a worm-hole. The quantum quench of the coupled chains is mapped to dynamics of the worm-hole. In the end we discuss the general procedure of applying this approach to interacting systems, and other open questions.
\end{abstract}
\maketitle

\tableofcontents

\section{Introduction}

In recent years, holographic duality, also known as anti-de-Sitter space/conformal field theory (AdS/CFT) correspondence\cite{maldacena1998,witten1998,witten1998b,gubser1998}, has attracted tremendous research interest in both high energy and condensed matter physics. This correspondence is defined as a duality between a $d+1$-dimensional conformal field theory defined on flat space and a $d+2$-dimensional quantum gravity theory defined on an AdS space background. In the known examples, the large-$N$ limit of the conformal field theory corresponds to the classical limit of the dual gravity theory. A key reason of such a correspondence is that the conformal symmetry group of $d+2$-dimensional space (with Lorentz metric) is ${\rm SO(d,2)}$, which is identical to the isometry group of AdS space. This duality can be generalized to more general field theories without conformal symmetry, which are dual to bulk gravity theories on different space-time manifolds.

\begin{figure}[htbp]
\centerline{\includegraphics[width=3.5in]{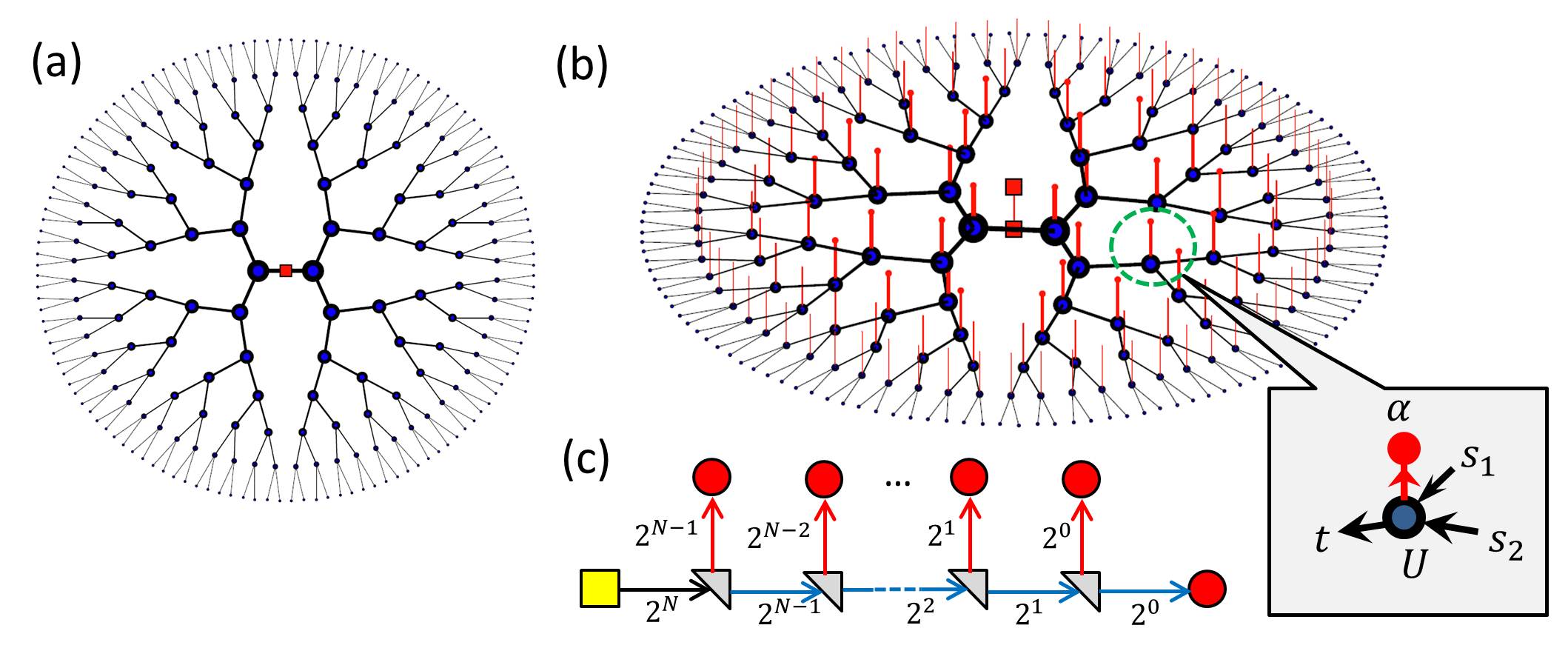}}
\caption{(a) Network representing a MERA state (without disentanglers). (b) Network of the exact holographic mapping. Each node in the network stands for a unitary transformation $U$ which maps the states of two input sites $\ket{s_1,s_2}$ to one bulk state (red dot) labeled by $\ket{\alpha}$ and one auxiliary state $\ket{t}$. More details of the definition are given in Sec. \ref{sec:def}.   (c) A simplified representation of the EHM network in (b), in which all the unitary transformations in the same layer are combined together to one unitary mapping (grey triangle). The boundary theory (yellow square) is mapped to auxiliary degrees of freedom (blue arrow) and bulk states (red filled circle) in each step. After $N$ steps, the boundary theory is mapped to the bulk theory consisting of all red dots. }\label{fig1}
\end{figure}

The holographic duality is intrinsically related to the renormalization group (RG) flow of the boundary theory\cite{akhmedov1998,boer2000,skenderis2002}, which is natural since the space-time dilatation is included in the conformal transformation group. The boundary flat space is mapped to the conformal boundary of the AdS space, and the emergent dimension perpendicular to the conformal boundary has the physical interpretation of energy scale. The RG flow of the boundary coupling constants become the bulk equation of motion\cite{heemskerk2011,lee2010,lee2011,lee2012}. Related to such ideas, B. Swingle\cite{swingle2012} has proposed a relation between holographic duality and multiscale entanglement renormalization ansatz (MERA)\cite{vidal2007,vidal2008}. MERA is a real space renormalization procedure defined for quantum states, which represents a highly entangled many-body state, such as the ground state of a conformal field theory, by a tensor network, as is illustrated in Fig. \ref{fig1} (a). Contraction of all tensors in this network defines an ansatz many-body wavefunction which can be used to approximate the ground state of the physical system. The network is viewed as a discretized version of the AdS space bulk theory\cite{swingle2012,swingle2012b,evenbly2011,hartman2013} , which is dual to the boundary CFT. This proposed correspondence provides a physical interpretation of the Ryu-Takayanagi formula\cite{ryu2006} which relates entanglement entropy to the minimal surface area in the bulk. The continuous generalization of MERA\cite{haegeman2013} and its relation to AdS/CFT\cite{nozaki2012} has also been discussed. However, there are important differences between MERA and AdS/CFT correspondence. In the former the bulk tensor network is classical even for a generic CFT, while in the later the bulk is only classical when the boundary theory is in the large $N$ limit. Although the network structure provides some information about the bulk geometry, this information is incomplete. In particular, the time direction metric is not explicitly encoded in the network, which makes it difficult to understand some interesting phenomena such as the correspondence between a finite temperature boundary system and a bulk black-hole geometry.

In this paper, we propose a generalization of MERA, the exact holographic mapping (EHM), which provides a more explicit and complete understanding of the bulk theory for a given boundary theory. Consider the network in Fig. \ref{fig1} (b), in which each vertex represents a unitary mapping which maps two input sites into two output sites. One of the two output sites (black) captures the low energy, or longer-range entangled degree of freedom of the two input sites, which becomes input of the next layers; The other output site (red) captures the high energy, or short-range entangled degree of freedom which is considered as a ``bulk" degree of freedom. The net effect of all layers of unitary transformations is a unitary mapping from the Hilbert space of the boundary theory to that of the bulk theory, defined by qubits living on the red sites. As has been discussed in Ref. \cite{vidal2008}, the MERA ansatz states are obtained by acting the reverse mapping (the one from bulk to boundary) on a direct product state in the bulk. Here we propose to apply the unitary mapping to all states in the boundary system, which leads to a bulk theory that is exactly equivalent to the boundary theory.
%The definition of EHM is illustrated in Fig. \ref{fig1} (b). The difference from MERA is that each isometry tensor which does coarse-graining in MERA is replaced by a unitary mapping which maps two input sites into two output sites. One of the two output sites (black) captures the low energy, or longer-range entangled degree of freedom of the two input sites, which becomes input of the next layers; The other output site (red) captures the high energy, or short-range entangled degree of freedom which is considered as a ``bulk" degree of freedom. The net effect of all layers of unitary transformations is a unitary mapping from the Hilbert space of the boundary theory to that of the bulk theory, defined by qubits living on the red sites. Since the mapping is unitary, the boundary theory is represented exactly by the bulk theory.
Properties of the boundary theory such as Hamiltonians and other operators, correlation functions and time-evolution can all be mapped to the bulk theory. %If we take a trivial bulk state with no entanglement between any two bulk sites, and act the inverse of the unitary mapping, the resulting boundary state has the form of a MERA ansatz state. In this sense, EHM generalizes MERA by allowing a more generic bulk state.
Compared with the AdS/CFT correspondence, we can consider the bulk theory obtained in EHM as a bulk ``matter field" living on the back-ground of hyperbolic space, while MERA corresponds to the infinite mass limit of the bulk matter field. (It should be noted that Ref. \cite{aguado2011} has mapped generic states of exactly solvable quantum double models using MERA, which can be considered as a realization of EHM. The MERA ansatz is exact in these models, meaning that all eigenstates of the boundary Hamiltonian correspond to direct product states in the bulk theory.)

Allowing the bulk matter field to have a nontrivial dynamics leads to important consequences in several different aspects. Firstly, this allows the bulk-boundary correspondence to be exactly defined for a generic boundary system, rather than for special ansatz states that can usually characterize the boundary physics only approximately. The entanglement in the bulk state, which is short-ranged for a properly chosen unitary mapping, can be viewed as a description of the ``residual entanglement" in the boundary theory that was not captured in the classical network itself. Secondly, the bulk state can be used to provide an independent definition of bulk geometry. By assuming the bulk state to be a massive state in the bulk, we can use the two-point correlation functions of the bulk state to define the distance between any pair of points in the bulk. When the EHM is chosen properly, such a definition of bulk distance can be interpreted as geodesic distance in a space-time manifold. In such a way, the bulk geometry corresponding to each given boundary system is not given by the network a priori, but is an ``emergent" property determined by the boundary state and the EHM. %More specifically, we propose to use the correlation functions of the bulk matter field to define the geodesic distance between bulk points. This definition allows us to determine the bulk geometry for each given boundary system.
In particular, since we have access to time-dependent correlation functions, we can not only probe the space geometry but also space-time geometry.

We illustrate the EHM by a $(1+1)$-dimensional lattice Dirac fermion example. We introduce an EHM for the free fermion and study the bulk space-time geometry for several different boundary states. For massless Dirac fermion at zero temperature, the corresponding bulk state can be considered as a discretized version of massive fermion on AdS$_{2+1}$. For massless fermion at finite temperature, the bulk state is consistent with a  Banados,
Teitelboim and Zanelli (BTZ) black-hole geometry\cite{banados1992}, although the same unitary mapping as the zero temperature case is used. The bulk sites in the center of the network (the ``infared" region) correspond to near-horizon region of the black-hole, and the entanglement entropy between bulk sites also provide a possible microscopic origin of the black-hole entropy. When the boundary state is a massive fermion at zero temperature, the corresponding bulk geometry is a space which effectively terminates at a certain radius, but by studying time direction correlation one can see that the ``infared boundary" of this geometry is not a horizon, and also the infared region does not carry entropy. As a more sophisticated example we study a quantum quench problem in which two coupled chains are entangled to form a massive ground state, and the coupling is turned off at a certain time. We show that the two entangled chains correspond to a bulk worm-hole geometry which connects to asymptotic AdS regions. After the quench, the worm-hole shrinks and then expands again, which can be compared with the holographic description of quantum quench procedure studied in the literature\cite{takayanagi2010,hartman2013}.

\section{Definition of the exact holographic mapping}\label{sec:def}

The exact holographic mapping is defined by multiplying a series of unitary transformations, as is illustrated in Fig. \ref{fig1} (b). We start from a lattice model with an $n$-dimensional Hilbert space on each site, and with total number of sites $L=2^N$. For two sites with states labeled by $\ket{s_1s_2},~s_{1,2}=1,2,...,n$, a unitary operator $U_{12}$ is defined to map them to two output sites labeled by $a$ (standing for ``auxiliary") and $b$ (``bulk").
\bea
U_{12}\ket{s_1s_2}=\sum_{t,\alpha=1,2,...,n}U_{s_1s_2}^{t\alpha}\ket{t}_a\otimes\ket{\alpha}_b
\eea
Here $\ket{\alpha}_b$ and $\ket{t}_a$ are sets of basis of the bulk site and auxiliary site, respectively.

The same transformation is carried for all pairs of sites $2i-1$ and $2i$, which leads to a unitary transformation on the Hilbert space of the whole system:
\bea
V_1=U_{12}\otimes U_{34}\otimes ...\otimes U_{2^N-1,2^N}
\eea
This is a mapping from a single chain with $2^N$ sites to two coupled chains, each with $2^{N-1}$ sites. Then we can define $V_2$ in the same way for the Hilbert space of $2^{N-1}$ auxiliary sites, which defines $2^{N-2}$ bulk sites and $2^{N-2}$ auxiliary sites in the second layer. Iterating this procedure $N$ times, as is illustrated in Fig. \ref{fig1} (c), we obtain a unitary mapping
\bea
M=V_NV_{N-1}...V_1
\eea
which maps the $2^N$ boundary sites to the same number of bulk sites. (In the last step, we have one bulk site and one auxiliary site, but both needs to be viewed as bulk sites.) Here $V_2$ should be understood as $V_2\otimes \mathbb{I}$ where $V_2$ acts on the auxiliary states in the first layer and $\mathbb{I}$ is the identity operator acting on the bulk states of the first layer. The $V_n$ in each layer should be understood in the same way.

For each choice of $U$ and a many-body state of the 1D chain $\ket{\Phi}$, the unitary operator $M$ maps $\ket{\Phi}$ to a 2D many-body state $\ket{\Psi}=M\ket{\Phi}$ defined on the network shown in Fig. \ref{fig1} (a). This mapping is defined for each state in the Hilbert space of the 1D chain, so that all operators, such as the Hamiltonian and the thermal density matrix, can also be mapped to the bulk system. If we take a direct product state $\ket{0}=\prod_{{\bf x}}\ket{0}_{\bf x}$ in the bulk, with ${\bf x}$ labeling the bulk sites, the corresponding boundary state $\ket{\Psi}=M^{-1}\ket{0}$ is an ansatz state defined in ordinary MERA approach. The mapping can be easily generalized by adding disentanglers\cite{vidal2007,vidal2008} in the same way as in MERA (which are unitary transformations on auxiliary sites that do not create new bulk sites), and/or by allowing $U$ to be different at different sites. Different from MERA case, such modifications are not necessary for characterizing the boundary state, since the mapping is exact. Therefore in this paper, we will focus on the simple choice described above with the same $U$ at each vertex, which already leads to rich consequences.

A key property of EHM that we will study is that the nontrivial bulk state $\ket{\Psi}=M\ket{\Phi}$ obtained from the mapping provides a measure of the bulk geometry. The belief behind this ``geometrical" point of view is that by appropriate choice of $U$, the bulk system can be gapped even if the boundary system is gapless. We do not have a proof of this statement for a generic boundary system, but this conjecture is supported by the fact that even a gapped bulk state on this hyperbolic geometry can provide the sufficient quantum entanglement that is necessary to characterize the boundary critical system. This is similar to the MERA case discussed in Ref. \cite{evenbly2011,swingle2012}.

With the assumption that we have mapped the boundary state to a massive bulk state, we can define the geodesic distance $d_{({\bf x},t_1),({\bf y},t_2)}$ between two bulk space-time points by the two point correlation function. For a massive state the two point function at long distance has the asymptotic form
\bea
\avg{O_{\bf x}(t_1)O_{\bf y}(t_2)}\simeq C_0 \exp\kd{-\frac{d_{({\bf x},t_1),({\bf y},t_2)}}{\xi}}
\eea
We use this equation as a {\it definition} of the distance function:
\bea
d_{({\bf x},t_1),({\bf y},t_2)}=-\xi\log\frac{\avg{O_{\bf x}(t_1)O_{\bf y}(t_2)}}{C_0}\label{distancedefgeneral}
\eea
The correlation length $\xi$ and the constant $C_0$ depend on the operator chosen, so that this equation can be used to determine the distance up to a constant and an overall scale. We have omitted the possible power law term front multiplying the exponential decay term, since the $\log$ of the correlation function will be dominated by the linear in $d$ term in the long distance limit.

Apparently, one would like to define the distance in a way that is independent from the choice of operator $O$. The suitable choice will be an appropriate upper bound of all two-point correlation functions. For equal time correlation function, there is an obvious choice of such a bound, which is the mutual information defined as
\bea
I_{{\bf xy}}=S_{{\bf x}}+S_{{\bf y}}-S_{\bf xy}
\eea
Here $S_{\bf xy}=-{\rm Tr}\kc{\rho_{\bf xy}\log\rho_{\bf xy}}$ is the von Neumann entropy of sites ${\bf xy}$ with reduced density matrix $\rho_{\bf xy}$, and similarly $S_{{\bf x}({\bf y})}$ is the entropy of a single site ${\bf x}({\bf y})$. Therefore the spatial distance can be defined by
\bea
d_{{\bf x}{\bf y}}=-\xi \log\frac{I_{{\bf xy}}}{I_0}\label{distancedef}
\eea
When each site ${\bf x}$ has $D$ states, the entropy $S_{\bf x}\leq \log D$. Therefore we have $I_{\bf xy}\leq 2\log D$. If two sites have $I_{\bf xy}=2\log D$, it means they are maximally entangled with each other and not entangled with any other sites. According to Eq. (\ref{distancedef}) ${\bf x}$ and ${\bf y}$ have minimal distance in this case. Therefore it is natural to define the distance between such a maximally entangled pair to be $0$, which means $I_0=2\log D$.

The geodesic distance was also related to mutual information in Ref. \cite{raamsdonk2010}, but the mutual information discussed there was between different regions in the boundary system. It is also interesting to note that the distance definition (\ref{distancedef}) may be related to the idea discussed recently in Ref. \cite{maldacena2013} that nonlocal quantum entanglement creates wormhole between far away spatial regions.

\section{Free fermion example}\label{sec:freefermion}

As an explicit example, we consider the $1+1D$ lattice Dirac fermion with the following Hamiltonian:
\bea
H=\sum_k c_{k}^\dagger\left[ \sigma_x\sin k+\kc{m+B\kc{1-\cos k}}\sigma_y\right]c_k\label{HDirac}
\eea
Here $\sigma_x,\sigma_y$ are Pauli matrices, and the annihilation operator $c_k$ is a two component spinor. In the long wavelength limit $k\>0$, the single particle Hamiltonian is approximately $k\sigma_x+m\sigma_y$ which approaches the continuous Dirac model. Now consider the unitary transformation $U$ which is a single-particle basis transformation
\bea
\vect{a_{i1}\\b_{i1}}=\frac1{\sqrt{2}}\kc{\ba{cc}1&1\\-1&1\ea}\vect{c_{2i-1}\\c_{2i}}\label{holomapping}
\eea
The spin index is omitted, which is preserved in this transformation. This mapping preserves the quadratic nature of the Hamiltonian, and breaks translation symmetry by doubling the unit cell. The Hamiltonian in the transformed basis is
\bea
H&=&H_a+H_b+H_{\rm int}\nn\\
H_a&=&\frac12\sum_ka_{k1}^\dagger\left[ \sigma_x\sin k+\kc{2m+B\kc{1-\cos k}}\sigma_y\right]a_{k1}\nn\\
H_b&=&\frac12\sum_kb_{k1}^\dagger\left[ \sigma_x\sin k+B(3+\cos k)\sigma_y\right]b_{k1}\nn\\
H_{\rm int}&=&\frac12\sum_k\kc{a_{k1}^\dagger\kd{i(1-\cos k)\sigma_x-iB\sin k\sigma_y}b_{k1}+h.c.}\nn\\
\eea
For the critical point at $m=0$, we see that $H_a$ has the same form as the original Hamiltonian except for a rescaling of the bandwidth by $\frac 12$. Since $H_a$ will be the input for the next layer, the low energy Hamiltonians of the auxiliary degrees of freedom for each layer are all related by a rescaling. The same holds for the bulk Hamiltonian $H_b$. In this sense the Hamiltonian of low energy degrees of freedom $a_k$ is at the ``fix point" of the EHM defined by $U$. The transformation above is iterated by defining
\bea
\vect{a_{i,n+1}\\b_{i,n+1}}=\frac1{\sqrt{2}}\kc{\ba{cc}1&1\\-1&1\ea}\vect{a_{2i-1,n}\\a_{2i,n}}\label{holomapping2}
\eea
leads to a bulk Hamiltonian $H_b=\sum_{\bf x,y}b_{\bf x}^\dagger h_{\bf xy}b_{\bf y}$.\footnote{There is an additional site $a_{0N}$ in the last layer, as is shown in Fig. \ref{fig1} (b) and (c). In the translation invariant Hamiltonians, $a_{0N}$ decouples from the rest of the sites, but in more general systems one should include the coupling of $a_{0N}$ with $b_{\bf x}$. For simplicity in the following we will omit $a_{0N}$ and focus on translation invariant states. } To distinguish bulk and boundary we will use $i$ to label boundary sites and use ${\bf x}=(x,n)$ to label bulk sites. Here $n$ labels the layer index and $x$ labels the sites in each layer. The bulk operators are related to the boundary ones by a unitary transformation
\bea
b_{\bf x}=\sum_i\phi_i^*({\bf x})c_i
\eea
The detail expression of the matrix element $\phi_i^*({\bf x})$ is given in the appendix. The basis wavefunction $\phi_i^*({\bf x})$ is actually a known basis called Haar wavelets\cite{haar1910}. (It is interesting to note that wavelets have been applied in renormalization group, which might be considered as a classical analog of the EHM approach.\cite{battle1992,best2000})

To understand the properties of the bulk theory, we study the bulk correlation functions. For the free fermion system studied here, we can use Wick theorem to determine all correlation functions by the single-particle Green's function:
\bea
G_{{\bf xy}\alpha\beta}(\tau)&=&\avg{Tb_{{\bf x}\alpha}(\tau)b_{{\bf y}\beta}^\dagger(0)}\nn\\
&=&\sum_k\phi_k^*\kc{\bf x}\phi_k\kc{\bf y}\avg{Tc_{k\alpha}(\tau)c_{k\beta}^\dagger(0)}\nn\\\label{Greens}
\eea

\begin{figure}[htbp]
\centerline{\includegraphics[width=1.8in]{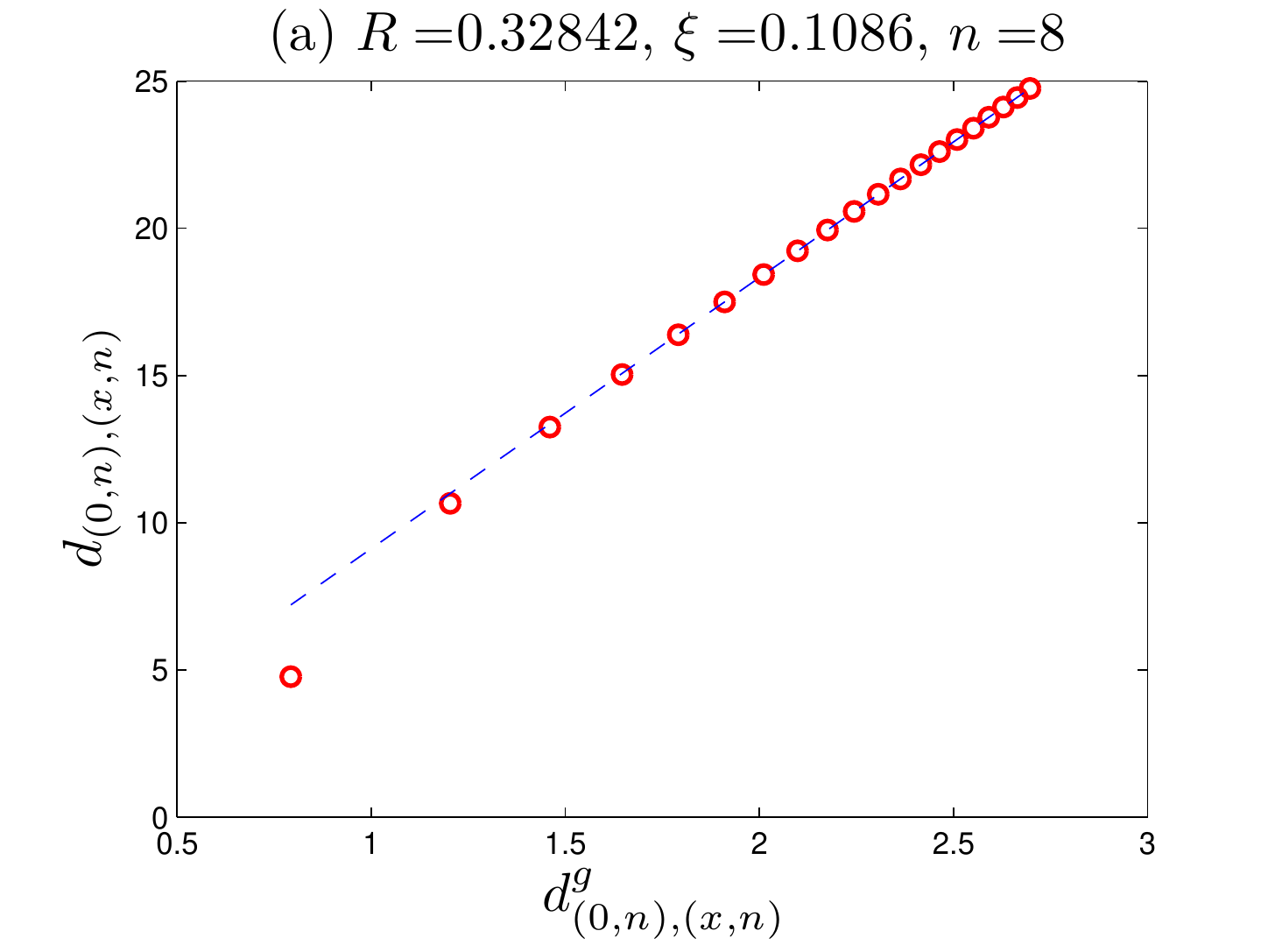}
\includegraphics[width=1.8in]{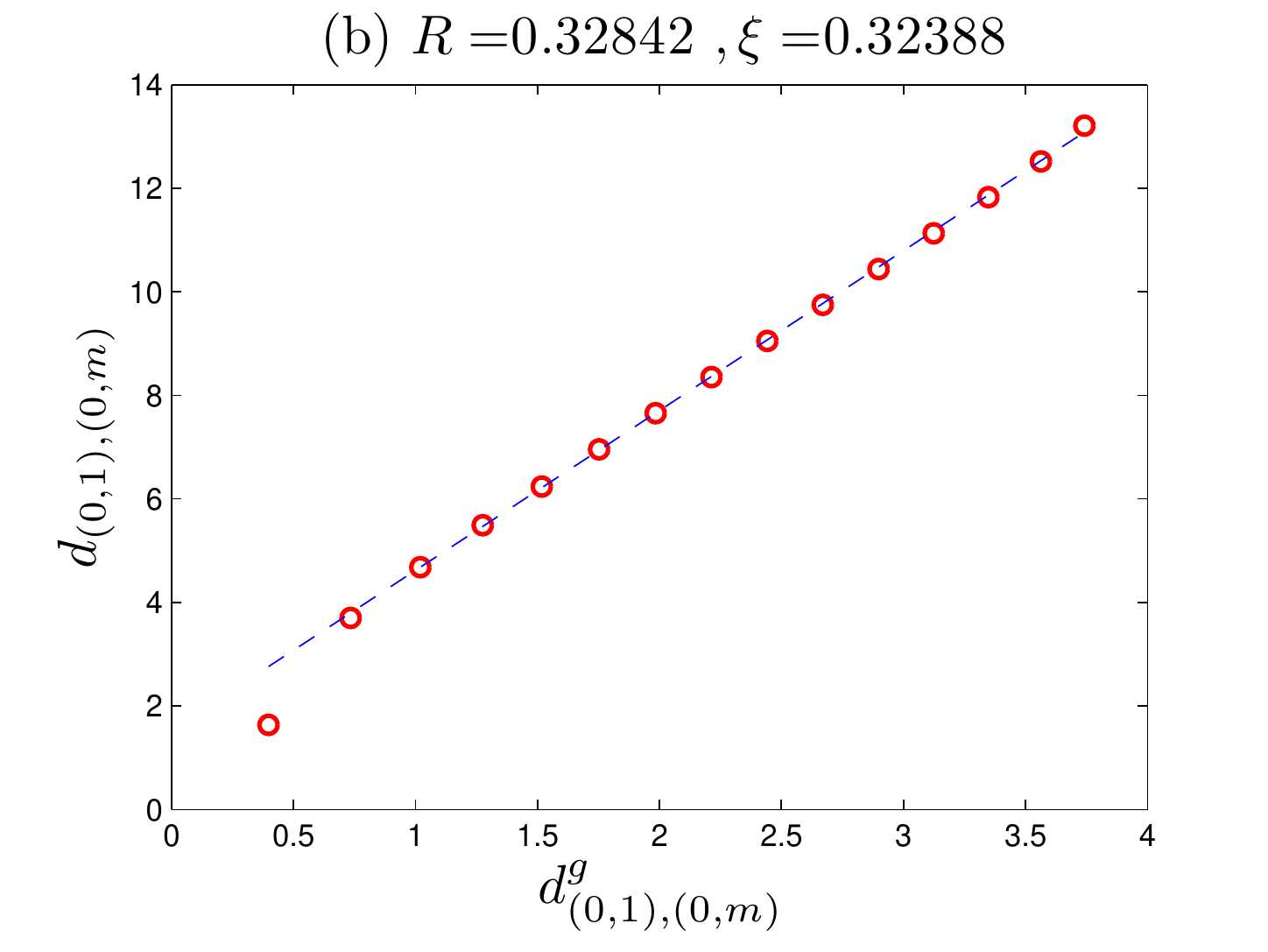}}
\centerline{\includegraphics[width=1.8in]{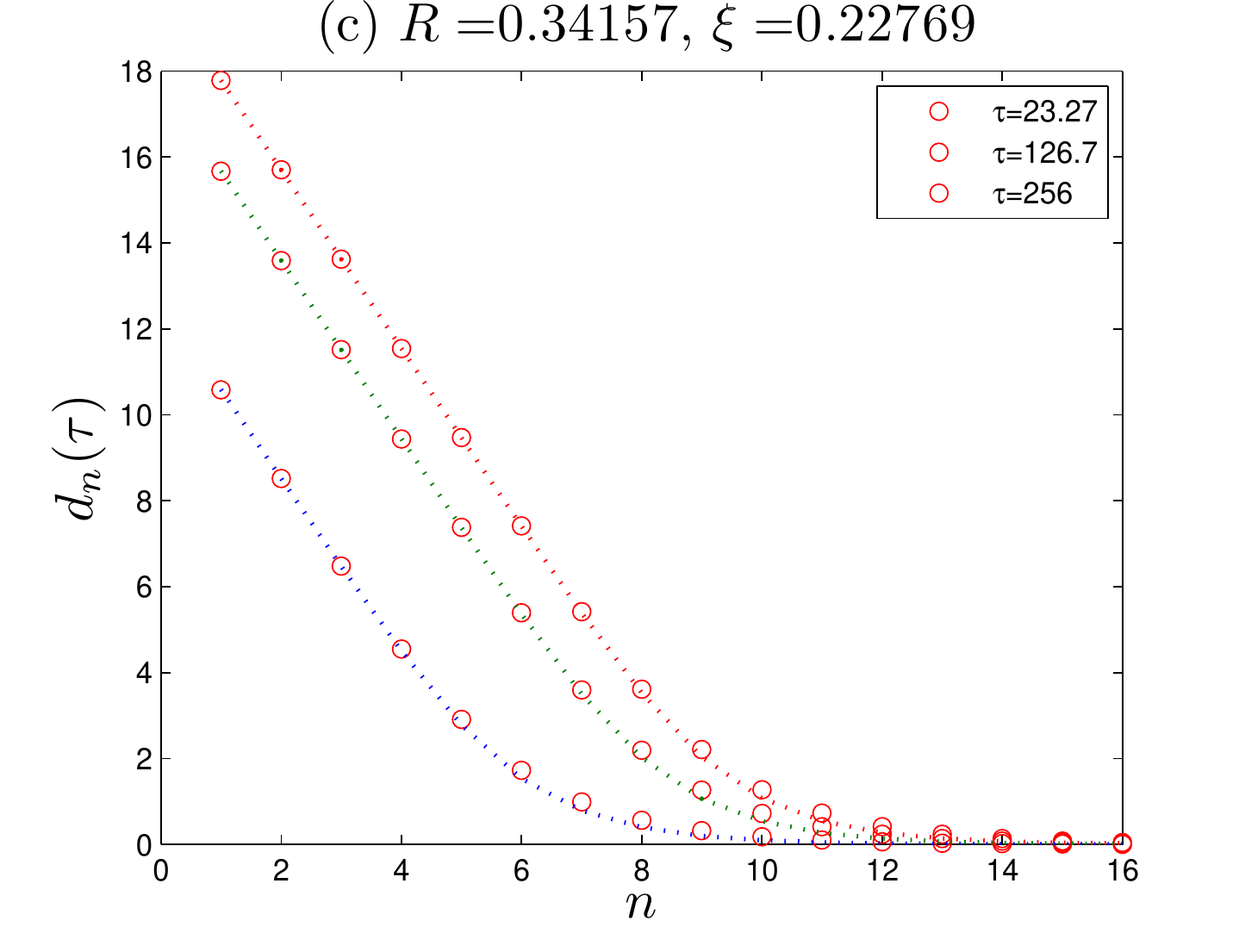}
\includegraphics[width=1.8in]{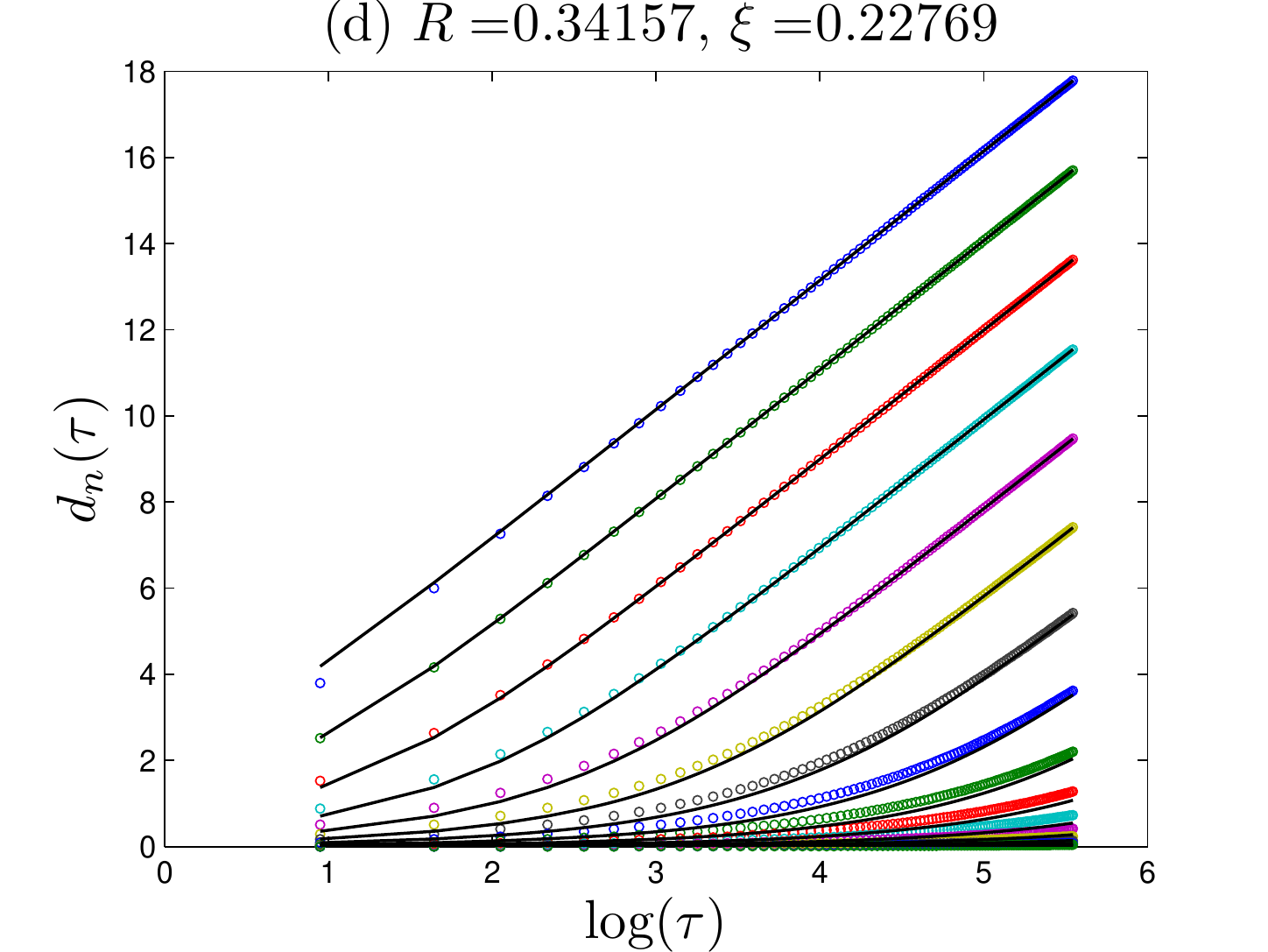}}
\caption{Distance between two points with equal time and different spatial location ((a) and (b)) and that between two points at the same spatial location with different time ((c) and (d)) for the critical system $T=m=0$. (a) and (b) shows the distance between two points separated along the horizontal direction and vertical direction, respectively. The $x$ axis is the geometrical distance between the two points in the AdS space. The AdS radius $R$ and correlation length $\xi$ is obtained from fitting in (a). (c) and (d) shows the distance between two points at the same spatial location and time difference of $\tau$. The fitting yields values of $R$ and $\xi$ independently from the spatial correlation functions. In all panels, the circles are the numerical results and the lines are the fitting with AdS space geodesic distance. All numerical calculations in Sec. \ref{sec:freefermion} and \ref{sec:finiteTm} are done for a chain with $2^17$ sites.}\label{MIcritical}
\end{figure}

We first study the spatial distance defined by mutual information in Eq. (\ref{distancedef}). The entropy $S_{\bf x}$ for the free fermion state is determined by the following formula\cite{peschel2003}:
\bea
S_{\bf x}=-{\rm tr}\kd{G_{\bf xx}\log G_{\bf xx}+\kc{\eye-G_{\bf xx}}\log\kc{\eye-G_{\bf xx}}}
\eea
with $G_{\bf xx}=\kd{G_{{\bf xx}\alpha\beta}(\tau\>0^+)}$ the equal time single particle correlation function matrix at site ${\bf x}$. The entropy of ${\bf y}$ and ${\bf xy}$ can be defined in the same way. Fig. \ref{MIcritical} (a) and (b) show the spatial distance between two points with the same $n$ and two points with the same $x$. We can see that the distance $d_{{\bf x}{\bf y}}$ scales like
\bea
d_{(x,n),(y,n)}\propto \log\abs{x-y},~d_{(x,n),(x,m)}\propto \abs{n-m}
\eea
This is consistent with the geodesic distance between two points in AdS space in the limit $d\gg R$.

To make closer comparison, consider the metric of Euclidean AdS$_{2+1}$
\bea
ds^2=\kc{\frac{\rho^2}{R^2}+1}dt^2+\frac{1}{\frac{\rho^2}{R^2}+1}d\rho^2+\rho^2d\theta^2
\eea
Here $\theta\in[0,2\pi)$ is an angle variable. To compare the discrete network with the AdS space, we notice that the perimeter of $n$-th layer is $2^{N-n}$, which should be identified with $2\pi \rho$. Therefore the point $(x,n)$ corresponds to
\bea
\rho=\frac{2^{N-n}}{2\pi},~\theta=\frac{x-2^{-1}+2^{-n-1}}{\rho}\label{coordinatedef}
\eea
 in the AdS coordinate. In the expression of $\theta$ we have introduced a constant shift such that the point in the $n+1$-th layer is in the middle of two sites in the $n$-th layer, as is shown in Fig. \ref{fig1}. The AdS geodesic distance is
\bea
d_{(x,n),(y,n)}&=&R{\rm arccosh}\kc{1+\frac{2\rho^2}{R^2}\sin^2\frac{\theta_1-\theta_2}2}\simeq 2R\log \frac{\abs{x-y}}R\nn\\
\label{horizontald}%\nn\\
%d_{(x,n),(x,m)}&=&R\kc{{\rm arcsinh}\rho_1-{\rm arcsinh}\rho_2}\simeq R\log 2\abs{n-m}
\eea
By fitting the formula (\ref{horizontald}) and (\ref{distancedef}) we can obtain $R\simeq 0.33$ and $\xi\simeq 0.11$, as is shown in Fig. \ref{MIcritical} (a). Using this value of $R$ we can compute the distance $d_{(0,1),(0,n)}$ between two points separated vertically. As is shown in Fig. \ref{MIcritical} (b), the mutual information between two sites $(0,1)$ and $(0,n)$ decays exponentially with their distance, although the slope gives a different correlation length $\xi$. The fact that $R<1$ tells us that the network can only characterize the large scale geometry of AdS space for length scale much larger than $R$.

In addition to the spatial geometry, we can also use time-ordered correlation functions to study the space-time geometry. In principle one should use a properly defined upper bound of all correlation functions for a given pair of space-time points. However, such a generalization of mutual information to space-time points is not known to us. Thus we instead consider the time-ordered single-particle Green's function
\bea
C_{\bf x}(\tau)=\sum_\sigma G_{{\bf xx}\sigma\sigma}(\tau),\label{TimeCor}
\eea
 with $G_{{\bf xy}\alpha\beta}$ defined in Eq. (\ref{Greens}). For simplicity we will use imaginary time. The time-direction distance is defined by the asymptotic behavior of $C_{\bf x}(\tau)$:
\bea
C_{\bf x}(\tau)=C_0e^{-d_{({\bf x},\tau),({\bf x},0)}/\xi}\label{distancedef2}
\eea
The distance defined by this equation can be compared with the geodesic distance in the AdS space
\bea
d_{({\bf x},\tau),({\bf x},0)}=R{\rm acosh}\left[\left(\frac{\rho^2}{R^2}+1\right)\cosh\frac {\tau}R-\frac{\rho^2}{R^2}\right]
\eea
Fitting of this formula can be used to independently determine the $R$ and $\xi$. As is shown in Fig. \ref{MIcritical} (c) and (d), the numerical results fit well with $R\simeq 0.34$ which is closed to the $R$ obtained from the spatial correlation function. However, $\xi$ is almost different by a factor of $2$. Such an anisotropy between space and time direction is a consequence of the difference between the two distance definitions (\ref{distancedef}) and (\ref{distancedef2}), and the different treatment of space and time in the EHM.

\section{Effect of finite temperature and finite mass}\label{sec:finiteTm}

A main advantage of EHM is that we can go beyond the scale invariant case and describe the space-time metric corresponding to a non-critical system without having to adjust the network itself. We can apply the same mapping $M$ to a different boundary system, which then leads to a bulk system with different correlation functions. If we still define the bulk geodesic distance using the correlation functions in Eq. (\ref{distancedef}) and (\ref{distancedef2}), we obtain a bulk geometry different from the AdS space. Two simplest ways to drive the system away from criticality are by adding a finite mass $m\neq 0$ and a finite temperature $T>0$.

\subsection{Finite $T$ system with $m=0$}

\begin{figure}[htbp]
\centerline{\includegraphics[width=1.8in]{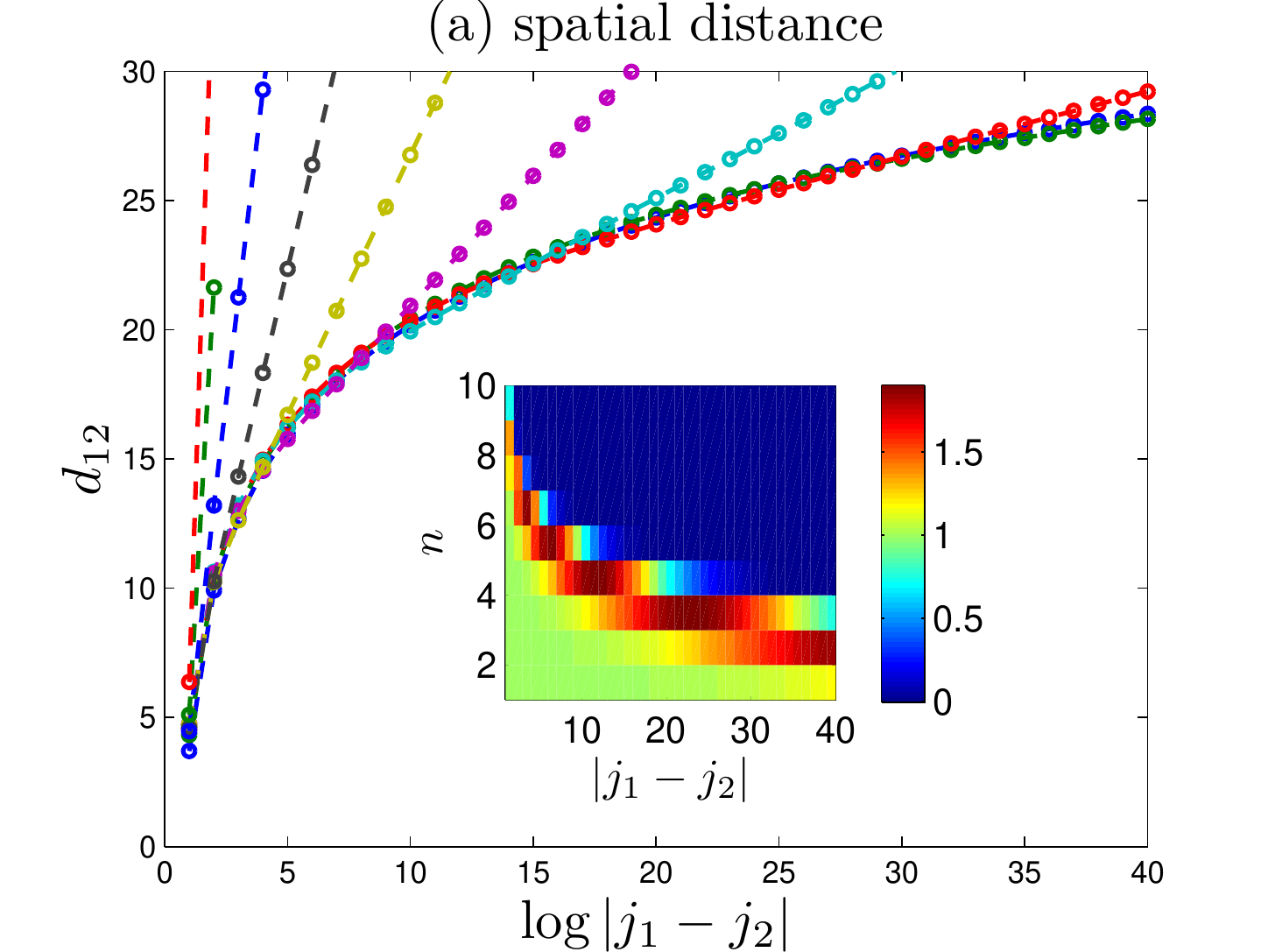}
\includegraphics[width=1.8in]{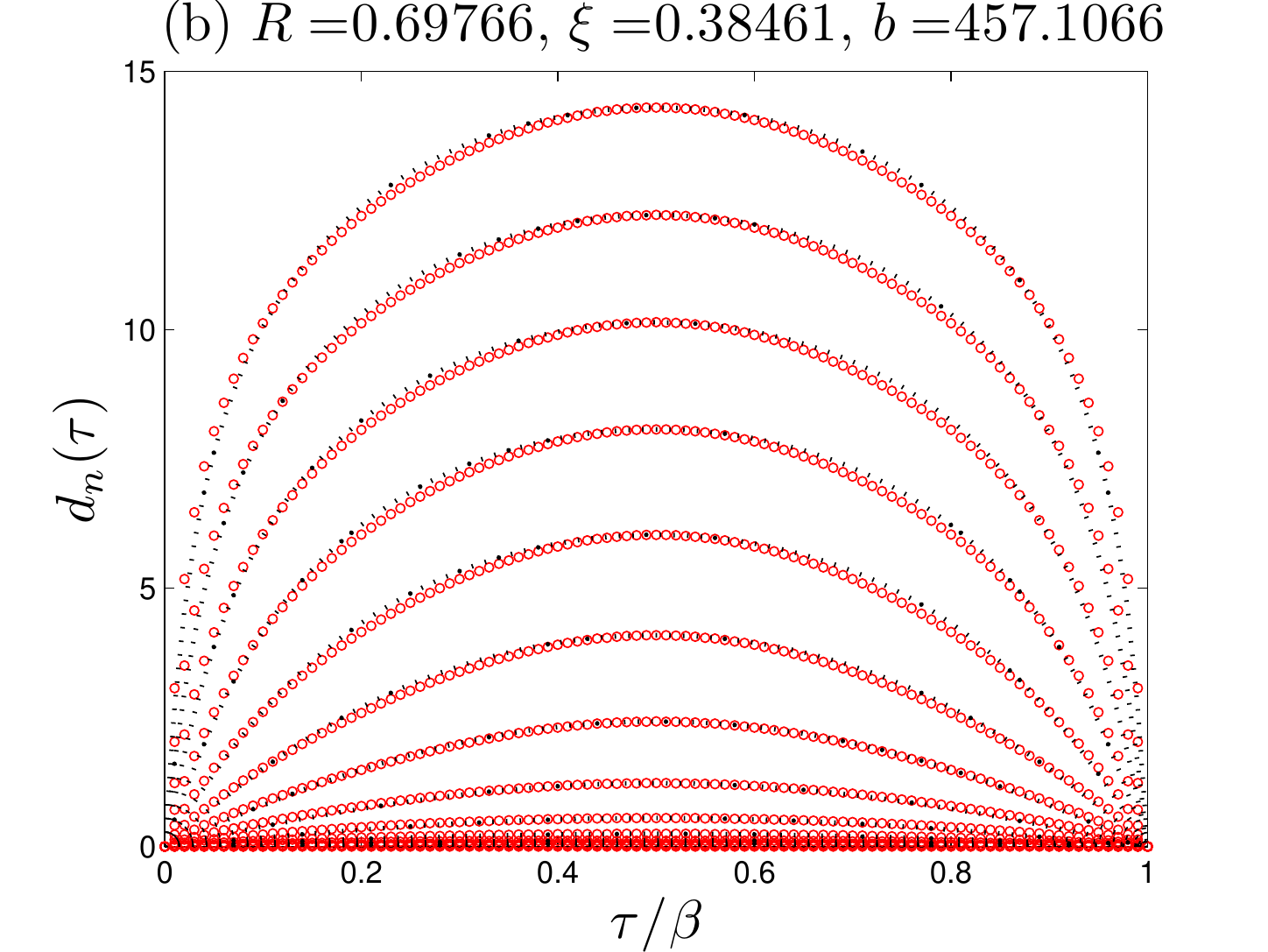}}
\centerline{\includegraphics[width=1.8in]{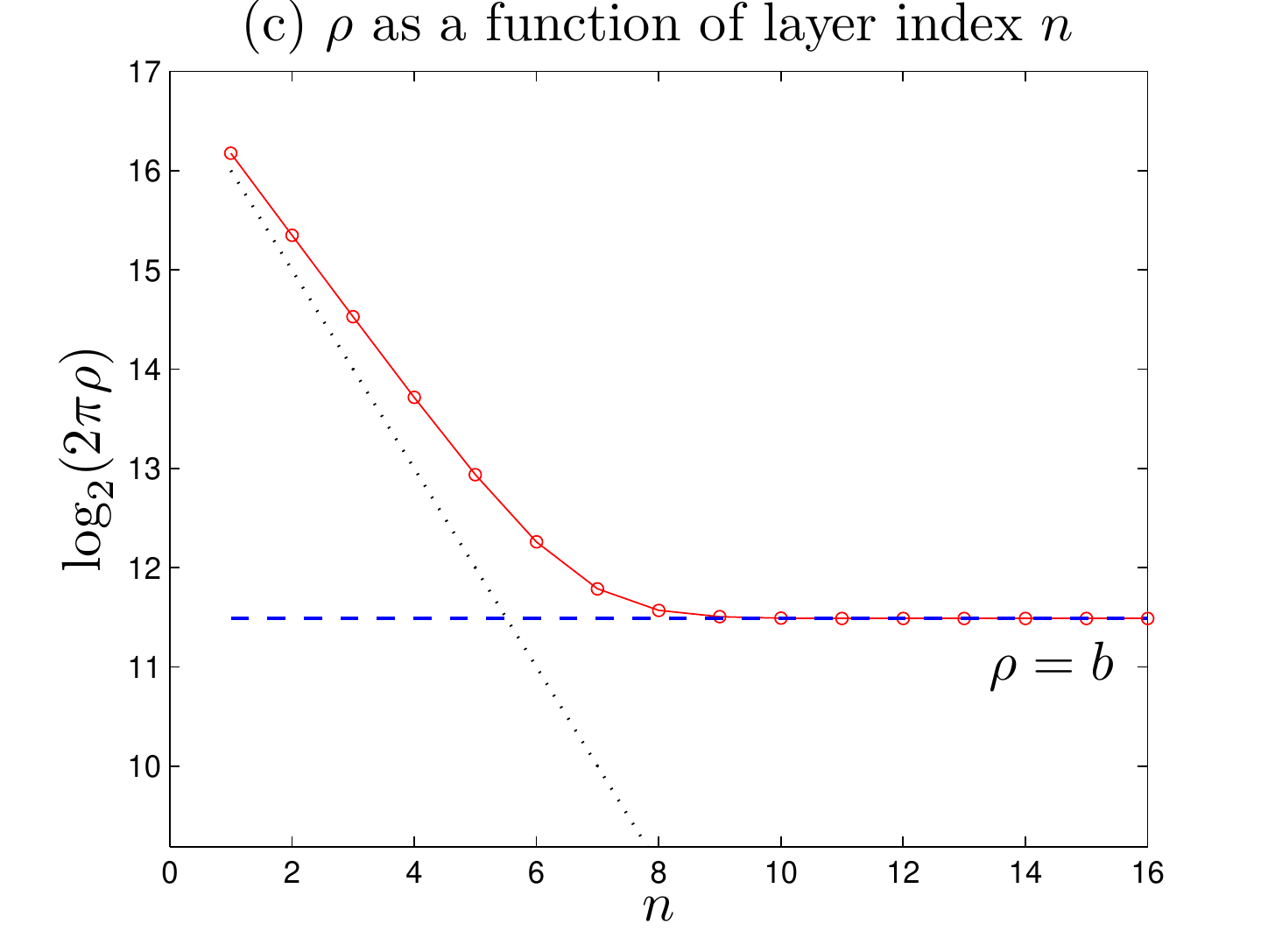}
\includegraphics[width=1.8in]{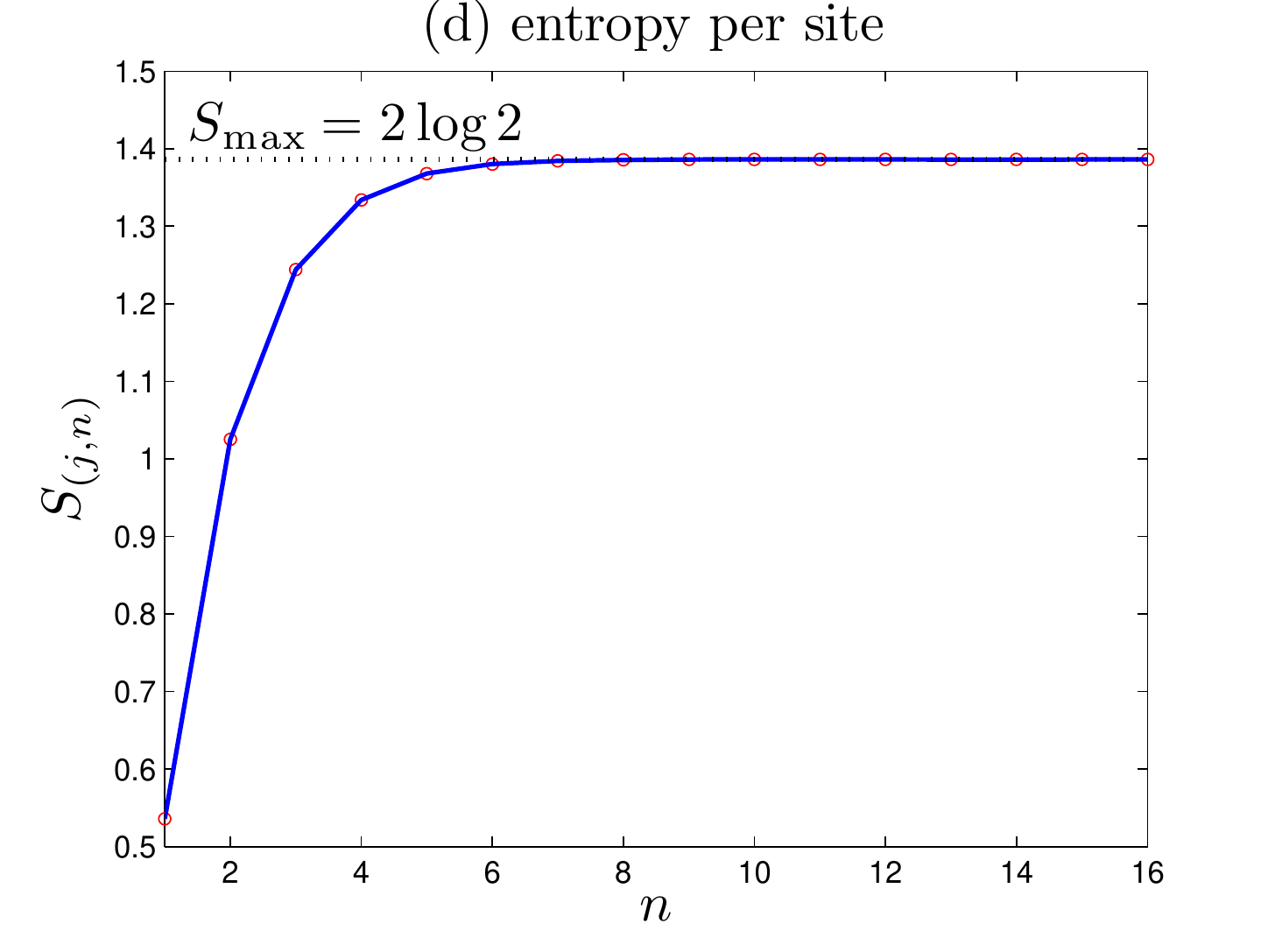}}
\caption{(a) Spatial distance $d_{12}$ between two points $(j_1,n)$ and $(j_2,n)$ for different $n$. Distance increases with the increase of $n$. The inset is the colorplot of the ratio of mutual information $I_{12}(T)$ of the finite temperature system and $I_{12}(0)$ that of the critical system, as a function of horizontal and vertical coordinates.  (b) Temporal distance $d_n(\tau)$ between two points $(j,n,0)$ and $(j,n,\tau)$ for different $n$. Distance decreases with the increase of $n$. The dashed lines are the fitting with the analytic formula (\ref{geodesicT}). (c) The radius $\rho$ as a function of $n$, obtained from the fitting (red line with circles). The blue dashed line labels $r=b$ and the black dotted line shows the zero temperature value $\rho=2^{N-n}/2\pi$. (d) Entropy per site as a function of $n$ for finite temperature (red circles) and zero temperature (blue line). The black dotted line shows the maximal entropy value $2\log 2$. All calculations are done for $T=0.005$. }\label{fig:MIfiniteT}
\end{figure}

We first study the system with $m=0$ and finite temperature $T>0$.  The space and (Euclidean) time direction distance are computed in the same way as zero temperature case, as is shown in Fig. \ref{fig:MIfiniteT}. In the spatial direction, the distance between two sites $(x,n)$ and $(y,n)$ as a function of the coordinate distance $|x-y|$ shows a cross over from $\propto\log\abs{x-y}$ (the zero temperature behavior) in short range to linear $\propto \abs{x-y}$ in long range, as is shown in Fig. \ref{fig:MIfiniteT} (a). The inset shows the ratio $I_{\bf xy}(T)/I_{\bf xy}(0)$ between the finite temperature and zero temperature mutual information, which shows a cross over from $1$ (green region) to $0$ (blue region). This is qualitatively consistent with the behavior of geodesic distance in a BTZ black hole geometry. The black-hole metric in Euclidean time is given by
\bea
ds^2=\frac{\rho^2-b^2}{R^2}d\tau^2+\frac{R^2}{\rho^2-b^2}d\rho^2+\rho^2d\phi^2
\eea
with $b$ the black hole radius. The distance between two points $(x,n)$ and $(y,n)$ at the same time is
\bea
d_{(x,n),(y,n)}=2R{\rm asinh}\kd{\frac{\rho}{b}\sinh\frac{b}{2R}},~\text{with}~\phi=\frac{2\pi \abs{x-y}}{2^{N-n}}
\eea
However, it is difficult to fit the numerical results with this formula because we cannot assume $\rho=2^{N-n}/2\pi$ any more. In the critical system, scale invariance determines that $\rho$ must scales in the same way as the lattice perimeter $2^{N-n}$, but for finite temperature $\rho$ is not determined {a priori}. To obtain $\rho$ we numerically calculate the distance in time direction $d_{({\bf x},\tau),({\bf x},0)}$ and fit it with the analytic geodesic distance
\bea
d_{({\bf x},\tau),({\bf x},0)}=R{\rm acosh}\kd{\frac{\rho^2}{b^2}-\kc{\frac{\rho^2}{b^2}-1}\cos\kc{\frac{2\pi}\beta\tau}}\label{geodesicT}
\eea
with $\beta=1/T$. From the fitting (Fig. \ref{fig:MIfiniteT} (b)) we obtain the parameters $R,\xi,b$ and also obtain the radius $\rho$ as a function of the vertical coordinate $n$, which is shown in Fig. \ref{fig:MIfiniteT} (c). Interestingly, $\rho$ approaches $b$ exponentially in the IR limit. Physically, this behavior reflects the fact that in IR limit (large $n$ limit) the bandwidth of the bulk states decays exponentially, so that the time-direction correlation length increases exponentially. In other words, the time direction correlation function decays more and more slowly in large $n$, as is shown in Fig. \ref{fig:MIfiniteT} (c).

The behavior of $\rho$ tells us that the IR region of the network now maps to the near-horizon region of the BTZ black hole geometry. This is an important difference from MERA case where a state with finite correlation length is simulated by a network truncated at finite depth\cite{evenbly2011}. An interesting relation to the black hole physics is given by studying the entanglement entropy of each bulk site with the rest of the system. Due to translation symmetry, $S_{(x,n)}=S_n$ is only a function of the vertical coordinate $n$. The entropy $S_n$ for both $T=0$ and $T=0.005$ is shown in Fig. \ref{fig:MIfiniteT} (d). From this result we see that the entropy per site quickly approaches the maximal value $2\log 2$, which means each bulk site, except those near the boundary, is maximally entangled with other sites, although the mutual information shows that the entanglement is only with nearby sites. It should be noticed that such a maximal entropy also shows that the state at the boundary is far from a MERA state defined by the same unitary mapping, as the latter with have a direct product bulk state with $S_n=0$ at each site. At finite $T$, the IR region has very long time-direction correlation length and very short spatial correlation length, which is interpreted as the neighborhood of black hole horizon. The maximal entropy carried by each site in this region can be considered as the origin of the black hole Bekenstein-Hawking entropy.

\subsection{Finite mass system with $T=0$}

\begin{figure}[htbp]
\centerline{\includegraphics[width=1.8in]{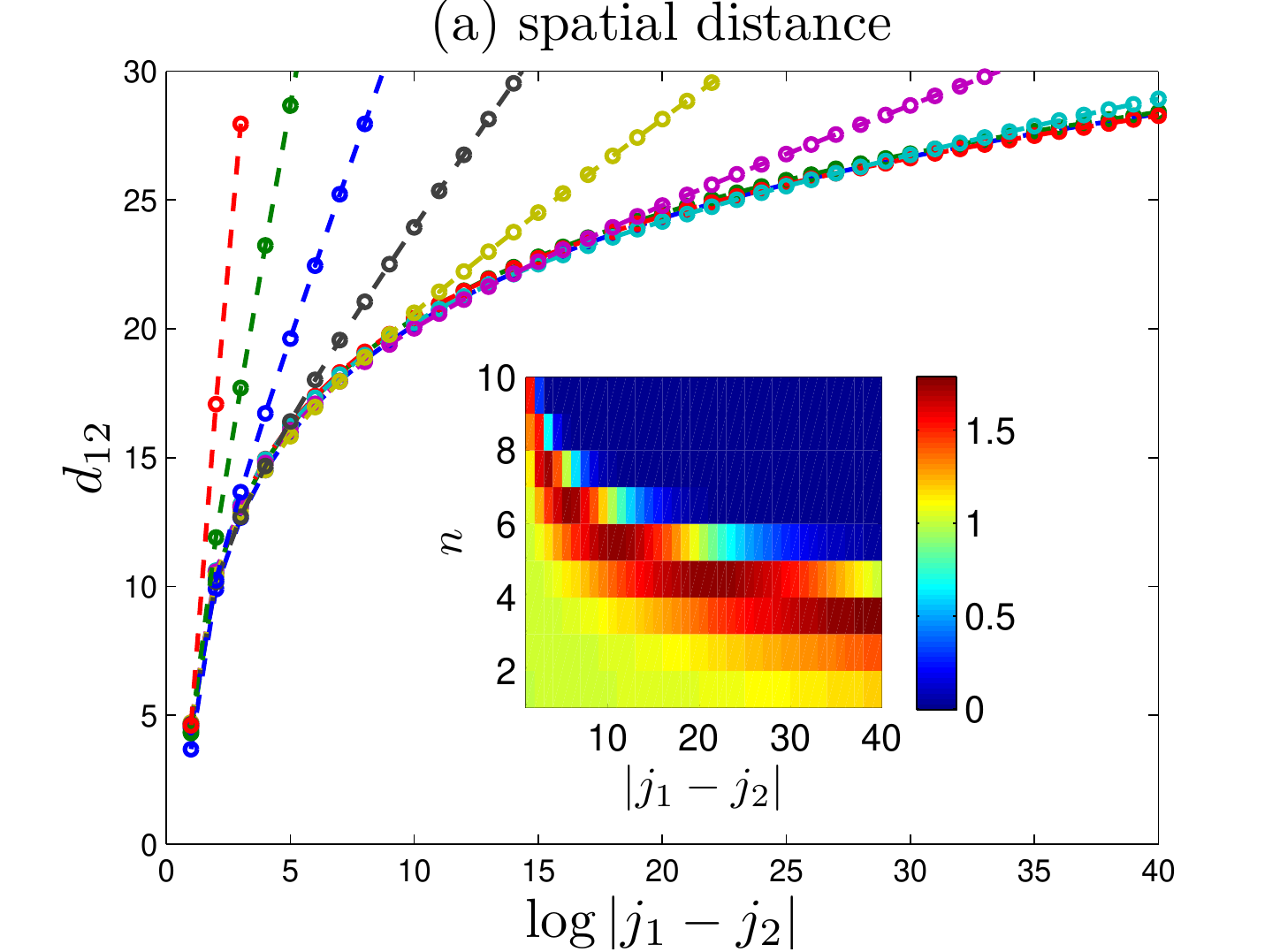}
\includegraphics[width=1.8in]{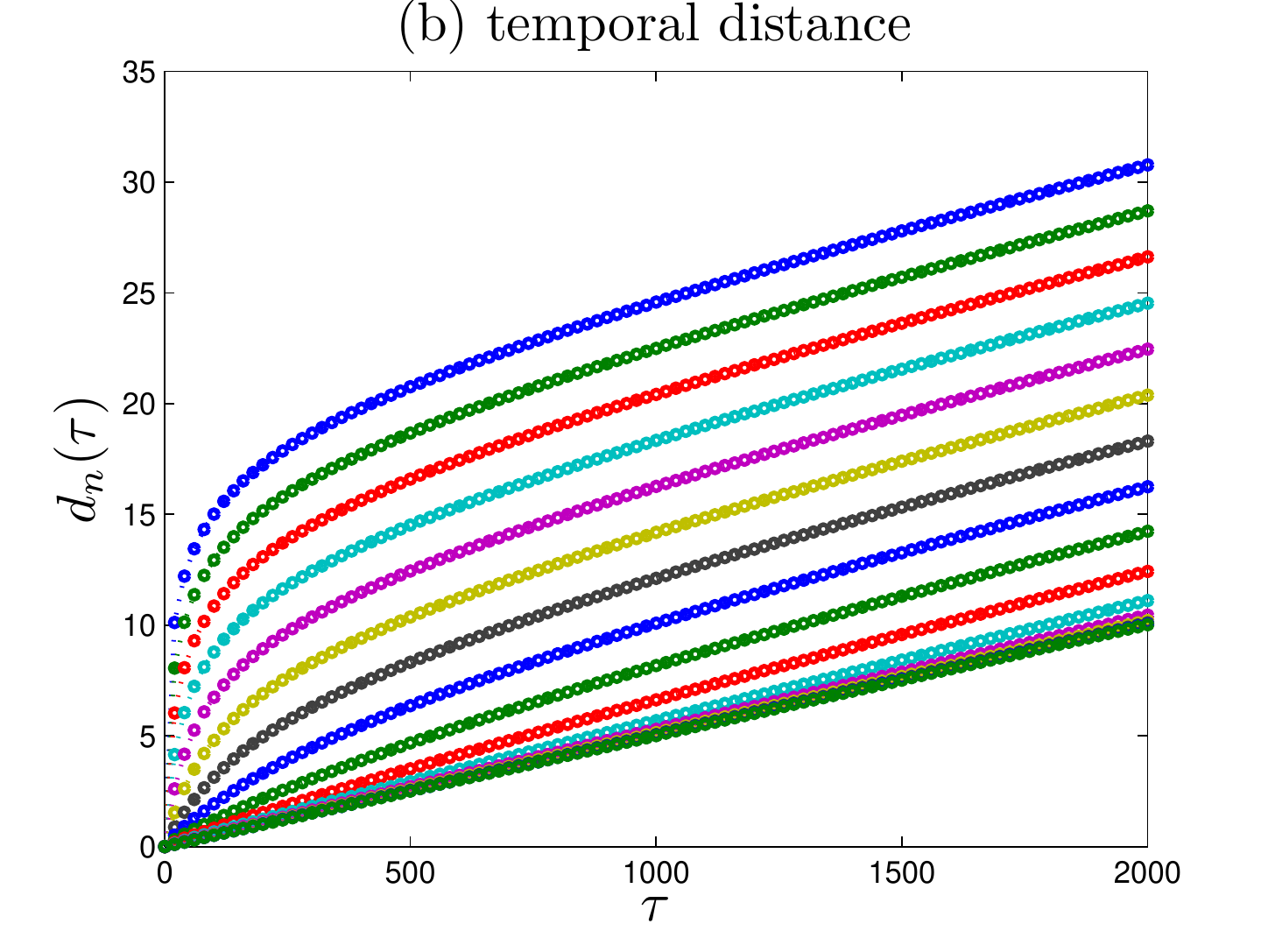}}
\centerline{\includegraphics[width=1.8in]{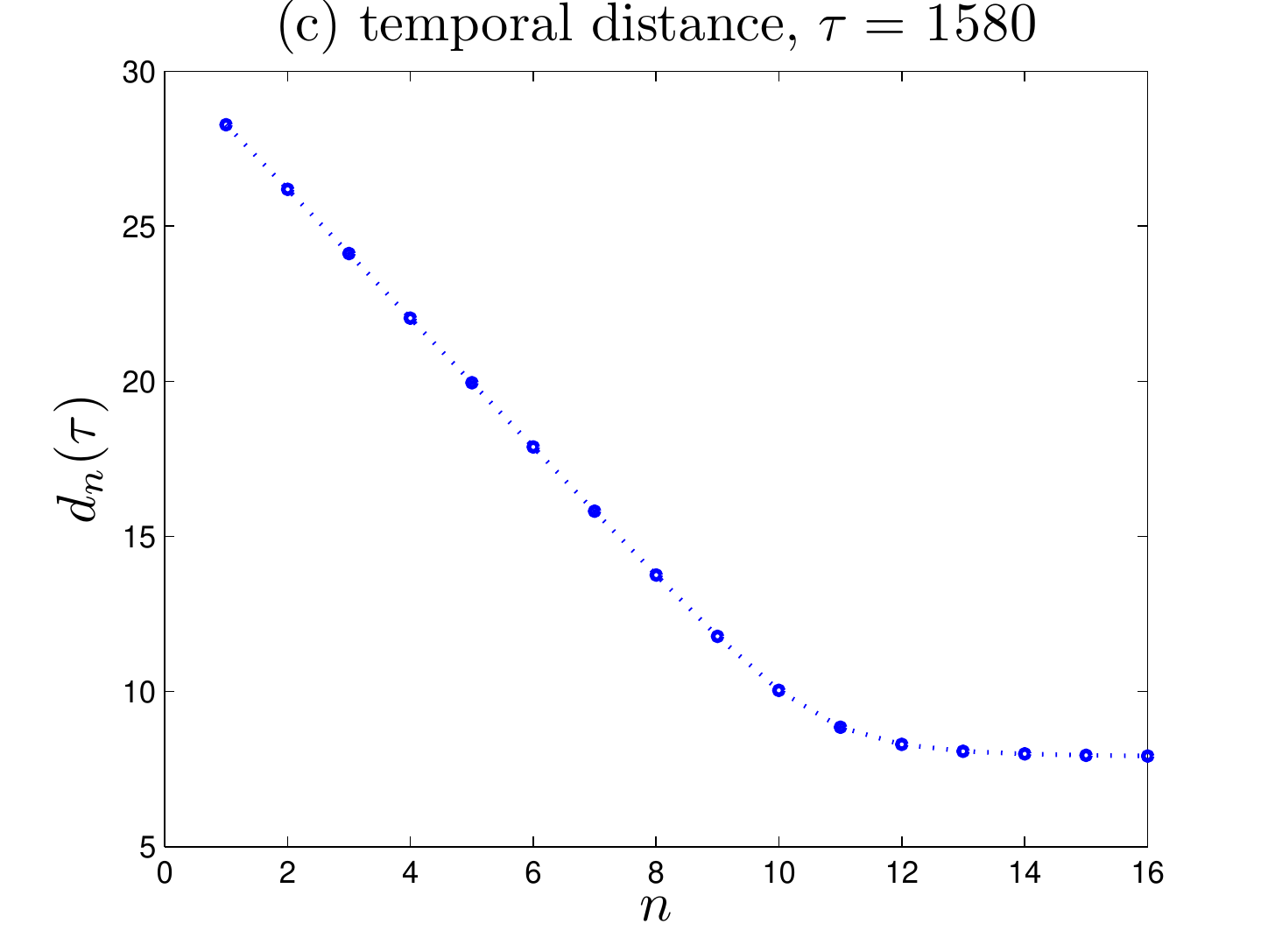}
\includegraphics[width=1.8in]{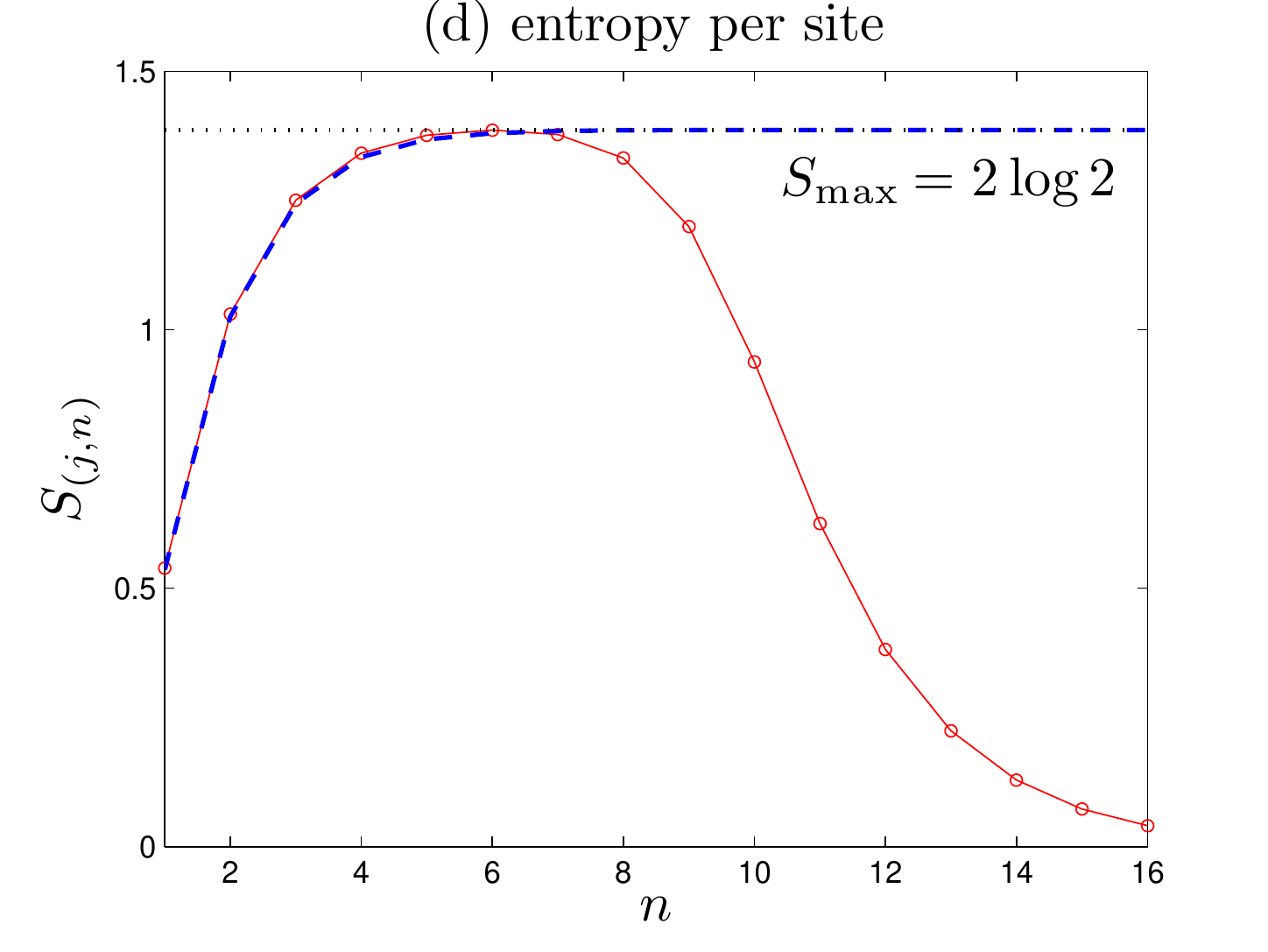}}
\caption{(a) Spatial distance $d_{12}$ between two points $(j_1,n)$ and $(j_2,n)$ for different $n$. Distance increases with the increase of $n$. The inset is the colorplot of the ratio of mutual information $I_{12}(m)$ of the massive system and $I_{12}(0)$ that of the critical system, as a function of horizontal and vertical coordinates.  (b) Temporal distance $d_n(\tau)$ between two points $(j,n,0)$ and $(j,n,\tau)$ for different $n$. Distance decreases with the increase of $n$. (c) Long time behavior of $d_n(\tau)$ as a function of $n$ for a $\tau=1580\gg 1/m$.  (d) Entropy per site as a function of $n$, for finite mass (red line with circles) and massless system (blue dashed line). The black dotted line show the maximal entropy value $2\log 2$. All calculations are done for $m=0.005, T=0$.}\label{fig:MIfinitem}
\end{figure}

Now we study the system with finite mass $m\neq 0$ and temperature $T=0$, which leads to a different bulk space-time with a characteristic scale given by $m$. As is shown in Fig. \ref{fig:MIfinitem} (a), the spatial distance behaves similarly as that of $T>0,~m=0$. In both cases, the spatial geometry has a IR cutoff scale, so that the distance between two points $(x,n)$ and $(y,n)$ interpolates from the $\log\abs{x-y}$ AdS behavior to the Euclidean $\abs{x-y}$ behavior. However, the time direction distance clearly distinguishes these two systems. As is shown in Fig. \ref{fig:MIfinitem} (b), in long time limit the distance $d_{({\bf x},\tau),({\bf x},0)}$ along the time direction increases linearly in $\tau$. This is simply a consequence of the exponential decay of correlation functions controlled by mass $m$. In the IR region, the spatial correlation becomes very short range, but the time-direction correlation length remains finite. Compared with the finite temperature case, one can see that this space-time has a spatial cut-off scale, but at the cut-off scale (``end of the space") the time direction remains finite. In other words, the different behavior of IR region for finite $T$ and finite $m$ shows that the IR boundary of the space-time is a light-like surface (the black hole horizon) in the former case, and a time-like surface in the latter case. Because the choice of space-time with such IR boundary is not unique, we have not fit the numerical results to a specific geometry. For example, one natural candidate metric to compare with will be the confined space-time proposed in Ref. \cite{witten1998}.

Besides the time-direction metric, another interesting difference between the finite mass and finite temperature systems is the behavior of bulk entanglement entropy. As shown in Fig. \ref{fig:MIfinitem} (d), the entropy of each bulk site $S_{\bf x}=S_n$ in UV region behaves similarly from the critical system (and the finite $T$ zero mass system), but in IR region the entropy is suppressed. Physically this is a consequence of the fact that the mass remains a constant during the mapping while the bandwidth decays exponentially in IR limit. In the geometric point of view, this is again consistent with the fact that the finite mass space-time terminates ``smoothly" and there is no entropy accumulated at the neighborhood of the cut-off scale, in contrast to the finite $T$ case.

\section{Worm-hole geometry and quantum quench process}

One advantage of the EHM approach is that it can be applied to generic boundary states, so that it can also characterize time-dependent processes. As an interesting example, we study the quantum quench process in two coupled chains, which is mapped to a ``worm-hole" geometry with two asymptotic AdS regions.

\begin{figure}[htbp]
\centerline{\includegraphics[width=1.8in]{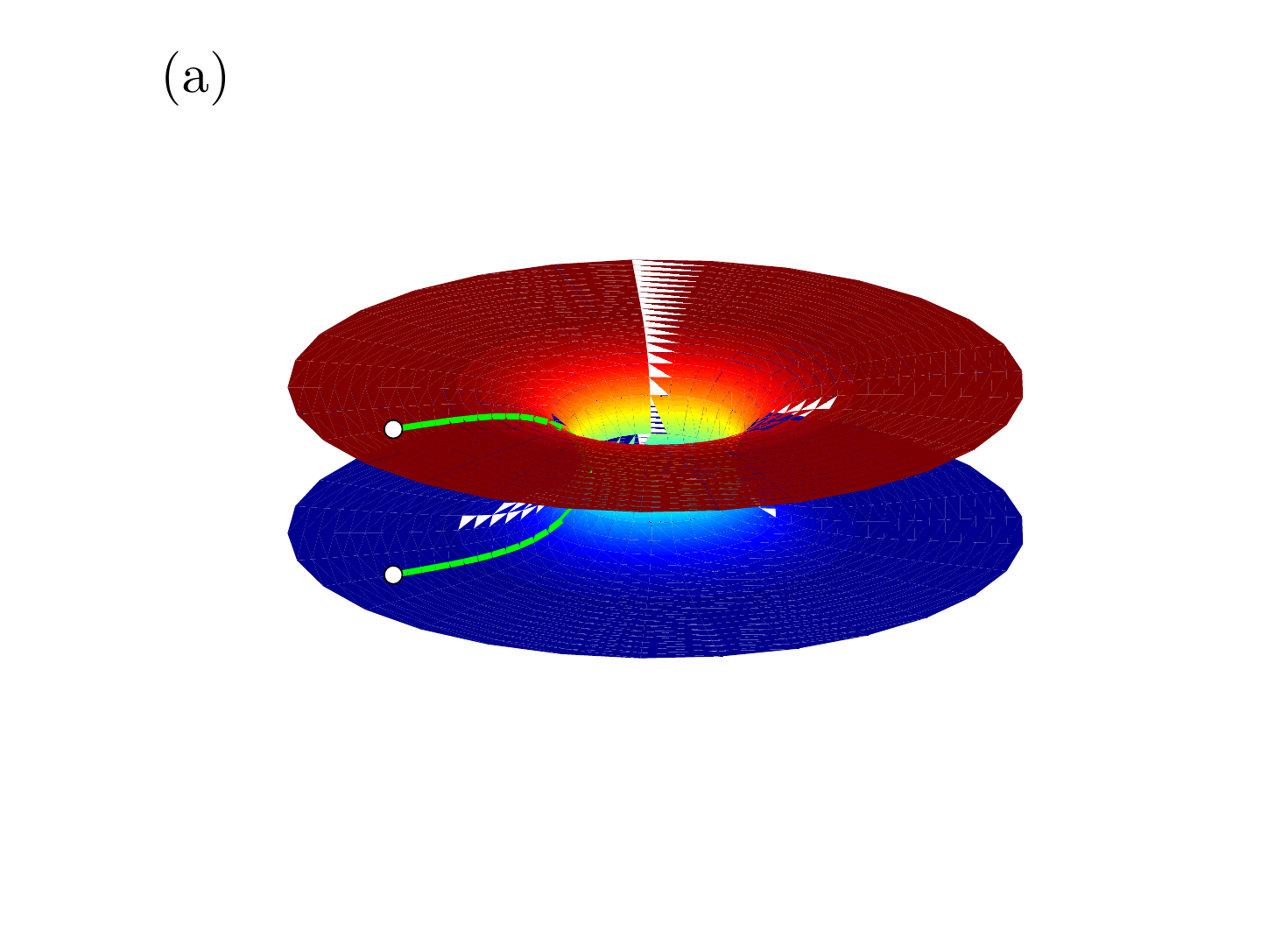}\includegraphics[width=1.8in]{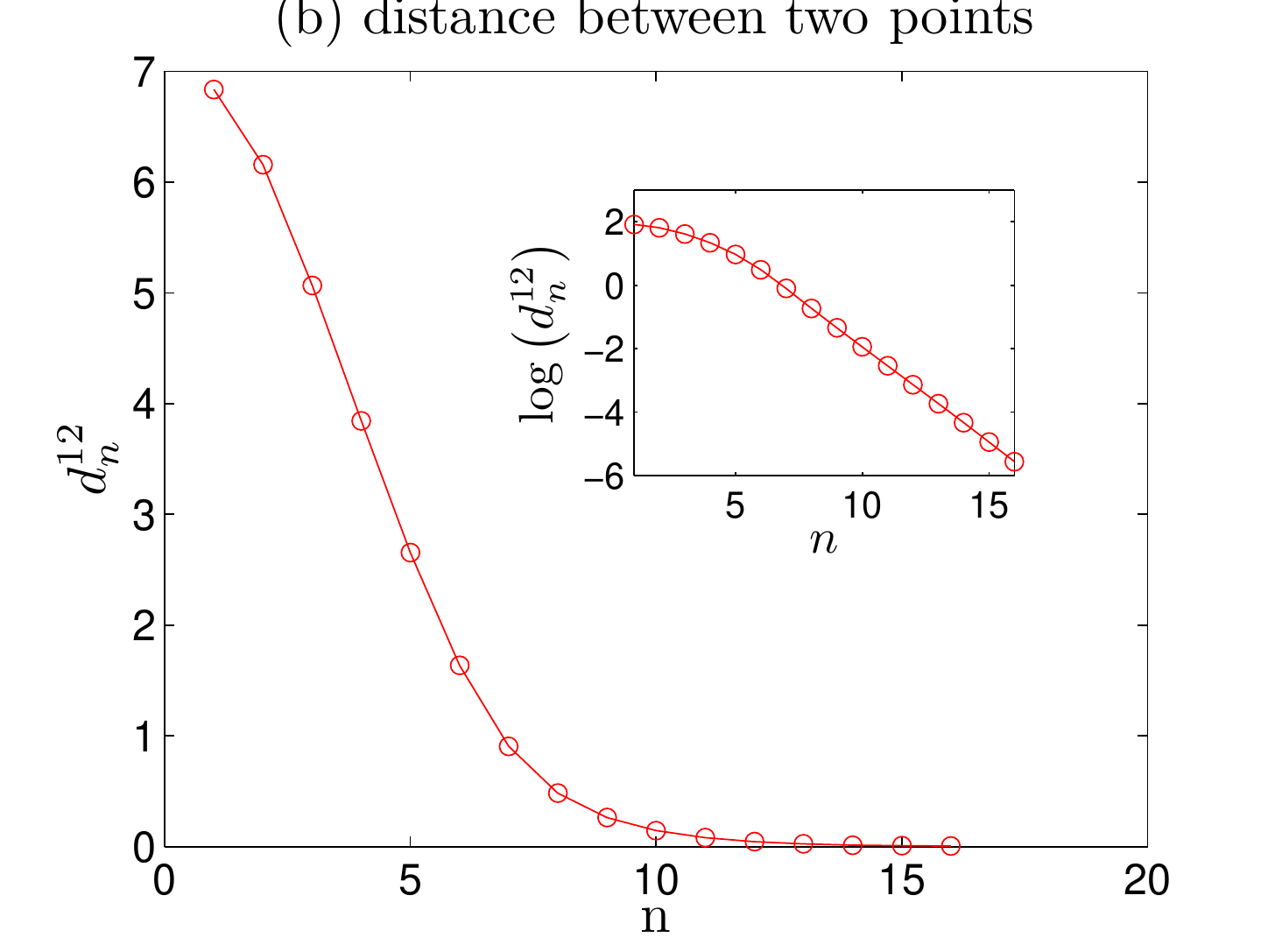}}
\caption{(a) Schematic picture of the wormhole geometry. The green line illustrates a geodesic path between the two end points. (b) Distance $d_{n}^{12}$ between two sites with coordinate $(j,n)$ in the two layers. The main figure and the inset shows the distance in linear scale and log scale, respectively. All the results in this section are done for $2^16$ sites and $\lambda=0.05$.  }\label{fig:bilayerMI}
\end{figure}

Consider the Hamiltonian
\bea
H&=&\sum_k\left(c_{k1}^\dagger~c_{k2}^\dagger\right)\kc{\ba{cc}h_k&\lambda \eye\\\lambda\eye&-h_k\ea}\vect{c_{k1}\\c_{k2}}
\eea
with $h_k=\sigma_x\sin k+B(1-\cos k)\sigma_y$ the Hamiltonian of a critical chain. The $\lambda$ term is a hopping that couples the two chains. We apply the holographic mapping (\ref{holomapping}) independently on the two chains. For $\lambda =0$ we will obtain two decoupled AdS-like spaces as we analyzed above. For $\lambda \neq 0$, entanglement occurs between the two chains in the ground state. Consequently, the mutual information between the two corresponding bulk spaces is non-vanishing. According to our definition of distance (\ref{distancedef}), this means that the distance between points in these two spaces is finite, {\it i.e.}, the bulk corresponding to the two coupled chains is now a connected topological space. To understand the bulk geometry we consider the distance between the sites $(j,n)$ in the two layers. The corresponding annihilation operators $b_{(j,n)}^{1,2}$ are superpositions of $c_{i1,2}$ correspondingly. The numerical results of the distance $d_{n}^{12}$ is shown in Fig. \ref{fig:bilayerMI}. Here we define $d_n^{12}=\log \frac{I_{\rm max}}{I_{(j,n)}^{12}}$ with $I_{(j,n)}^{12}$ the mutual information between the two sites at position $(j,n)$, and $I_{\rm max}$ the maximal possible mutual information between two sites $I_{\rm max}=2{\rm max}S_{\bf x}=4\log 2$. This choice of $I_{\rm max}$ means that we define the distance between two maximally entangled sites to be zero. From Fig. \ref{fig:bilayerMI} one can see that in the UV limit (small $n$) the distance between the two sites scales linearly with $n$, while in the IR limit the distance decays exponentially. The (exponentially) vanishing distance suggests that the two UV regions are connected by a worm-hole, similar to the one obtained from analytic continuation of a black-hole.\cite{israel1976}

\begin{figure}[htbp]
\centerline{\includegraphics[width=3.1in]{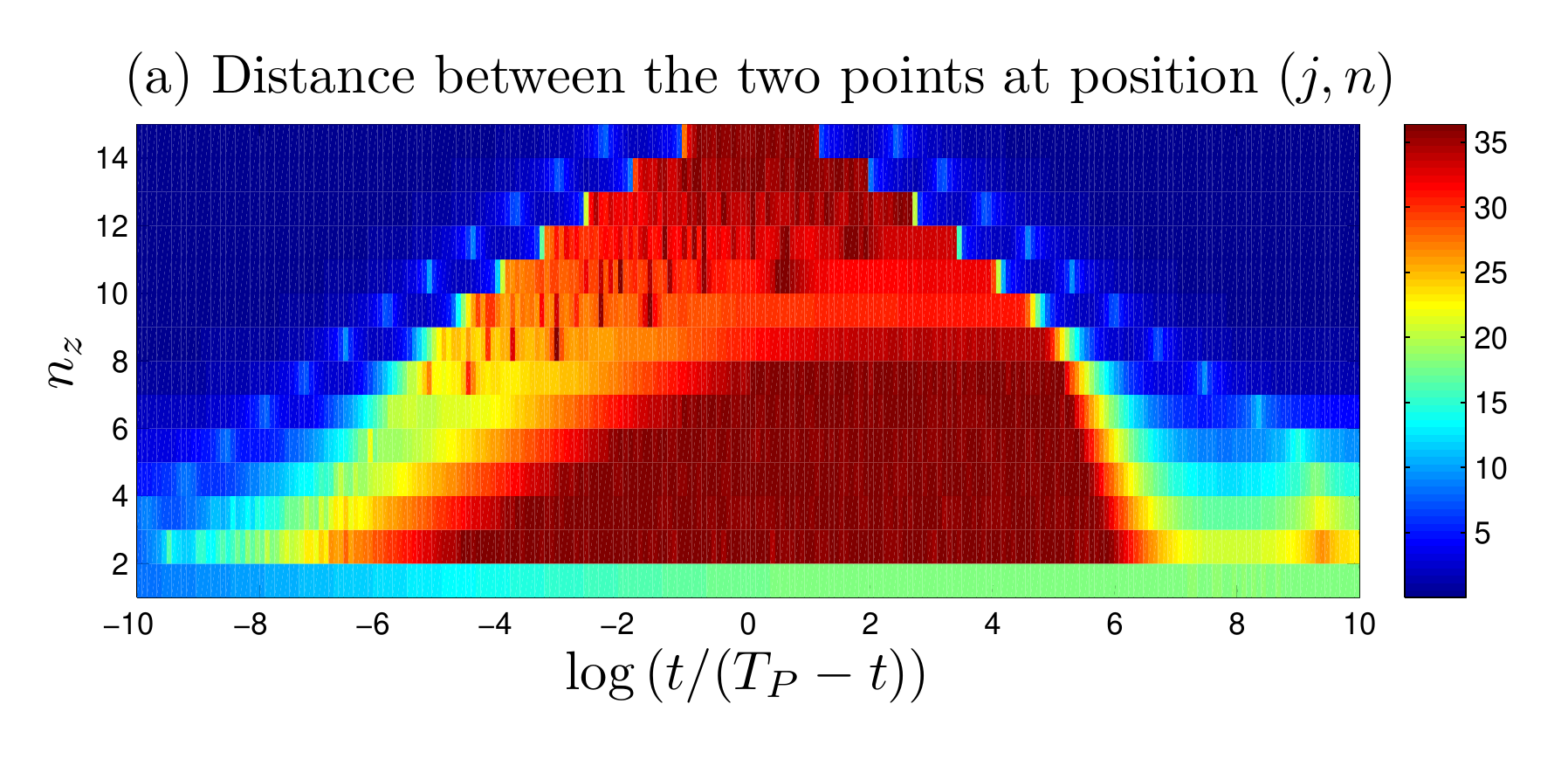}}
\centerline{\includegraphics[width=2.7in]{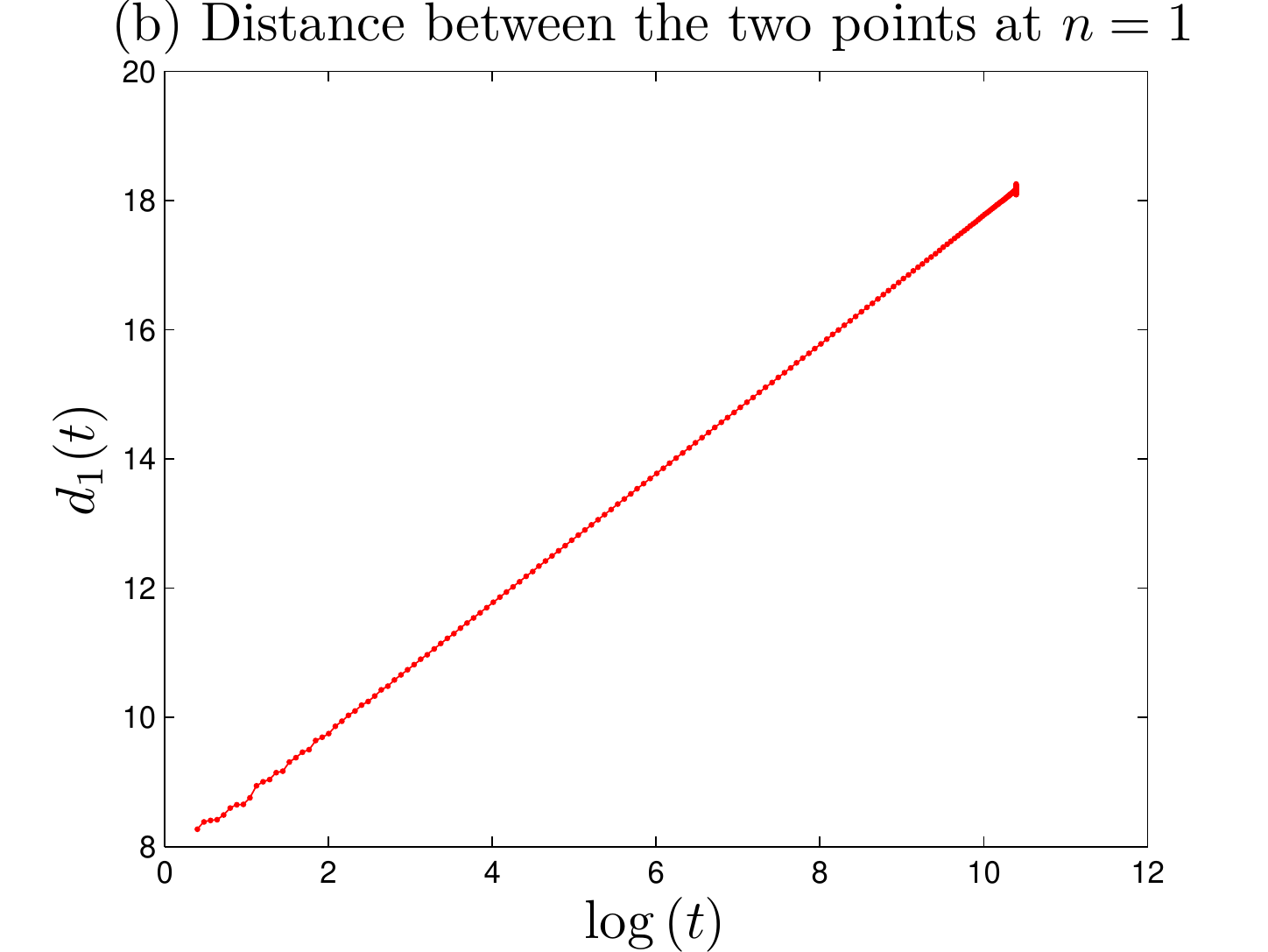}}
\centerline{\includegraphics[width=3.1in]{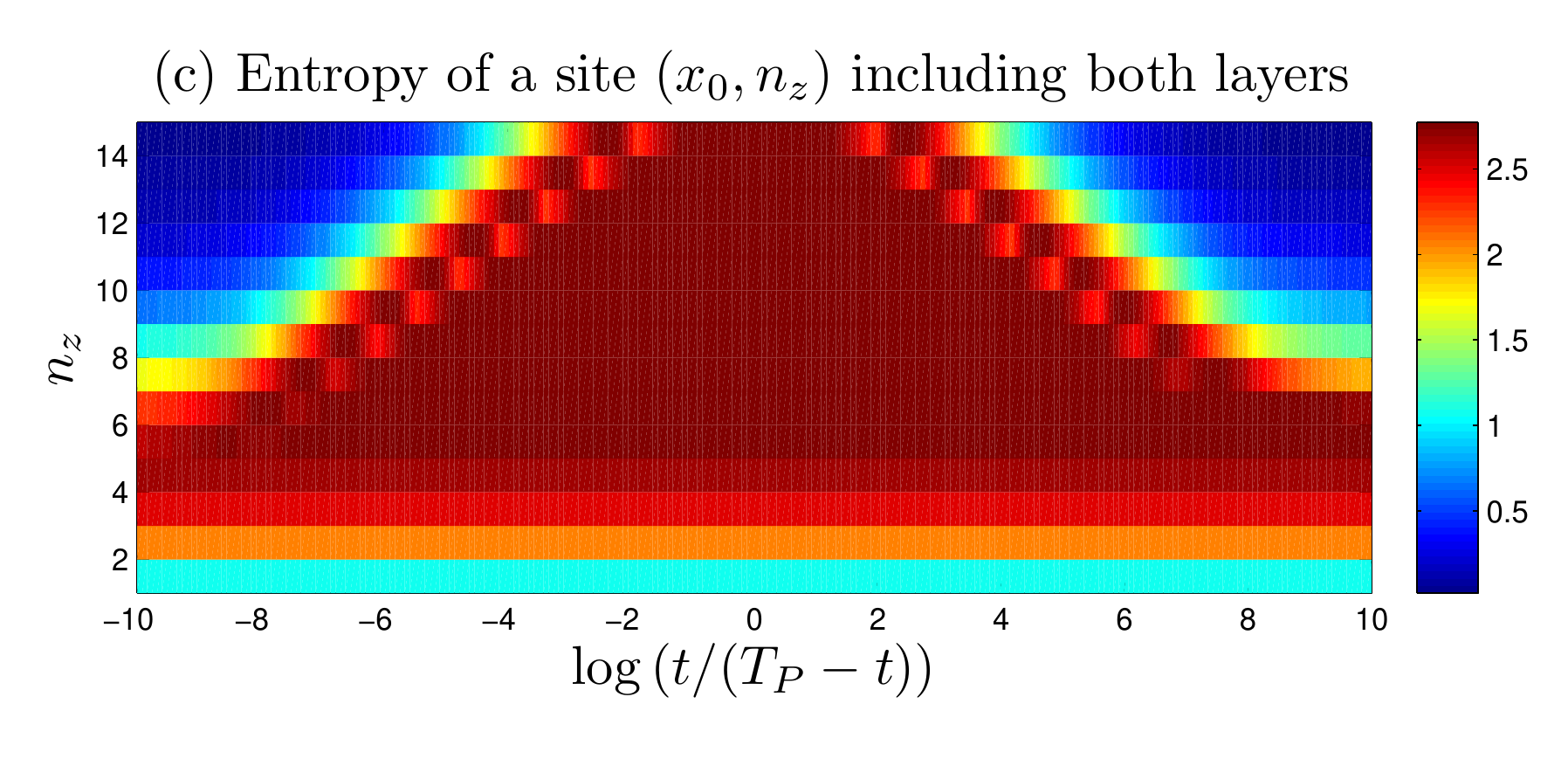}}
\caption{(a) Distance between two sites at position $(j,n)$ in the two layers as a function of radial coordinate $n$ and time. (b) The distance between two sites at the first layer $(j,1)$ as a function of $\log(t)$. (c) The entanglement entropy of the two sites at $(j,n)$ with the rest of the system as a function of radial coordinate $n$ and time. The blue region in (a) and (c0 is the wormhole where both the distance between the two sites and the net entropy of the two sites are exponentially small. }\label{fig:quench}
\end{figure}

In a recent work\cite{hartman2013}, the behavior of quantum entanglement is studied in a quantum quench problem with the wormhole geometry as the initial state. We can consider a corresponding quantum quench problem in our system by turning off the coupling $\lambda$ at time $0$. Before the quench, the system is at the ground state $\ket{G(\lambda)}$ of finite $\lambda$. After the quench, $\lambda=0$ and the time evolution of the two chains are independent. The single-particle Green's function can be obtained by
\bea
G_{{\bf xy}\alpha p,\beta q}\kc{t_1,t_2}&=&\sum_k\phi_k^*\kc{\bf x}\phi_k\kc{\bf y}\bra{G(\lambda)}c_{k\alpha p}(t_1)c_{k\beta q}^\dagger(t_2)\ket{G(\lambda)}\nn\\
c_{k\alpha p}(t_1)&=&\kd{e^{ih_kt_1}}_{\alpha\beta}c_{k\beta p}(0)\label{Greensdouble}
\eea
Here $p,q=1,2$ labels the layers and $\alpha,\beta$ labels the spin. It should be noticed that real time evolution instead of the imaginary Euclidean time is considered here. Using the single particle Green's function we can define the space-time metric by Eq. (\ref{distancedef}) and (\ref{distancedef2}) in the same way as the static case.

As an example of the correlation function we study the time-evolution of the distance between the two points at site ${\bf x}=(j,n)$. At time $t=0$ the distance $d_n^{12}(0)$ gives the worm-hole geometry shown in Fig. \ref{fig:bilayerMI}. The time evolution $d_n^{12}(t)$ is shown in Fig. \ref{fig:quench} (a). After the quench, the size of the worm-hole shrinks quickly and then expand again. The shrinking of the wormhole corresponds to a thermalization of the excitations created by the quench, and the reexpansion of the wormhole is a dethermalization procedure. In a generic system this should only occur in an Poincare recurrence time which is exponentially long in the system size, but in the free electron system it occurs in a time $T_P=L/2v$ with $L=2^N$ the system size and $v$ the speed of light of the system.\cite{takayanagi2010} In our model $v=1$ and $T_P=2^{N-1}$.

To see the time-evolution clearly, in Fig. \ref{fig:quench} we change the time variable $t$ to $f=\log\frac{t}{T_P-t}$. For $t\ll T_P$, $f\simeq \log \frac{t}{T_P}$, so that Fig. \ref{fig:quench} (a) tells us that the decrease of the wormhole size (blue region) is proportional to $\log t$. We also studied the distance between two points close to the boundary at $n=1$, which is also proportional to $\log t$. This result is different from the observation of Ref. \cite{hartman2013} where the geodesic distance between two boundary points increases linearly in $t$.\cite{maldacenaprivate} (The area of minimal surface connecting two boundary regions is studied there, and for AdS$_{2+1}$ the minimal surface reduces to the geodesic line.) Physically, the linear $t$ increase of distance in Ref. \cite{hartman2013} corresponds to an exponential decay of mutual information between the two points, while the $\log t$ dependence we obtain corresponds to a $1/t$ dependence of the mutual information. This difference is possibly because the following difference between free fermion theory and an interacting theory. In a free fermion system a single particle excitation propagates in space but remains a single particle, while in an interacting theory the particle can decay into multiple other particles. Therefore, for the free fermion theory the mutual information between the two bulk sites ``propagates" into a region with size $t$ (the speed of light is taken to be $1$). In other words, the mutual information remained is proportional to $1/t$. In contrast, in an interacting theory the mutual information can  ``propagate" to one of the many-body states in the region with size $t$. Since there are $D^t$ states in this region, with $D=4$ the number of states at each site, the remaining mutual information will be estimated by $D^{-t}$.

Such a difference provides an example when the geodesic distance we define by mutual information is inconsistent with the minimal surface area required by the Ryu-Takayanagi formula of entanglement entropy, although in the simpler cases of single chain, they qualitatively agree. This is probably related to the fact that the distance defined by Eq. (\ref{distancedef}) is generically different from the geometrical distance of the ``classical" network we use to define $M$. More discussion about this will be given in Sec. \ref{sec:generals}.

 Another quantity we calculate is the entanglement entropy of the two sites at $(j,n)$ with other sites in the bulk. At $t=0$, in the IR region the two sites at $(j,n)$ are almost maximally entangled with each other, and the net entropy of the two sites $S^{12}_{(j,n)}$ almost vanishes. After the quench, the entanglement starts to delocalize, and the net entropy of the two sites increases quickly. As is shown in Fig. \ref{fig:quench} (b),  entropy is filled into infared region when the wormhole shrinks. During the dethermalization period, the entropy is removed.

\section{Some more general analysis of EHM}

In the three sections above, we have restricted our discussions to EHM in free fermion systems, for which the bulk properties can be computed exactly. The EHM can in principle be applied to more generic interacting systems, but the bulk or boundary properties cannot be computed exactly for the general cases. However, one can still understand some generic properties of the EHM, which we will discuss in the following.

\subsection{Causal cone structure}

An important feature of the MERA ansatz state is the existence of a causal cone structure\cite{vidal2008,evenbly2013}. To compute the reduced density matrix of a boundary region for a MERA state (which determines all the physical variables in that region, such as energy average value), one does not need the information about the whole network, but only need the network in a region in the bulk, named as the causal cone. The causal cone only contains $\sim \log L$ number of sites when the boundary system has $L$ sites. The causal cone structure is essential for the efficient calculation of physical quantities in the MERA state.

\begin{figure*}[htbp]
\centerline{\includegraphics[width=5in]{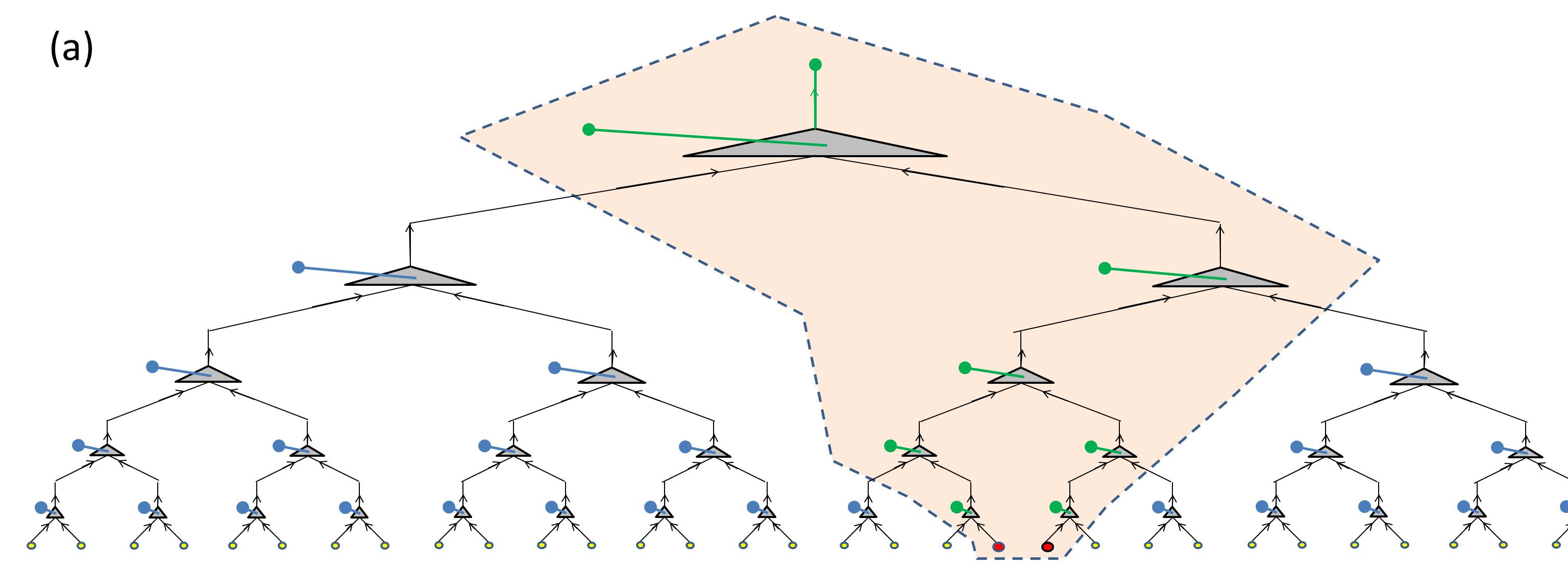}}
\centerline{\includegraphics[width=5in]{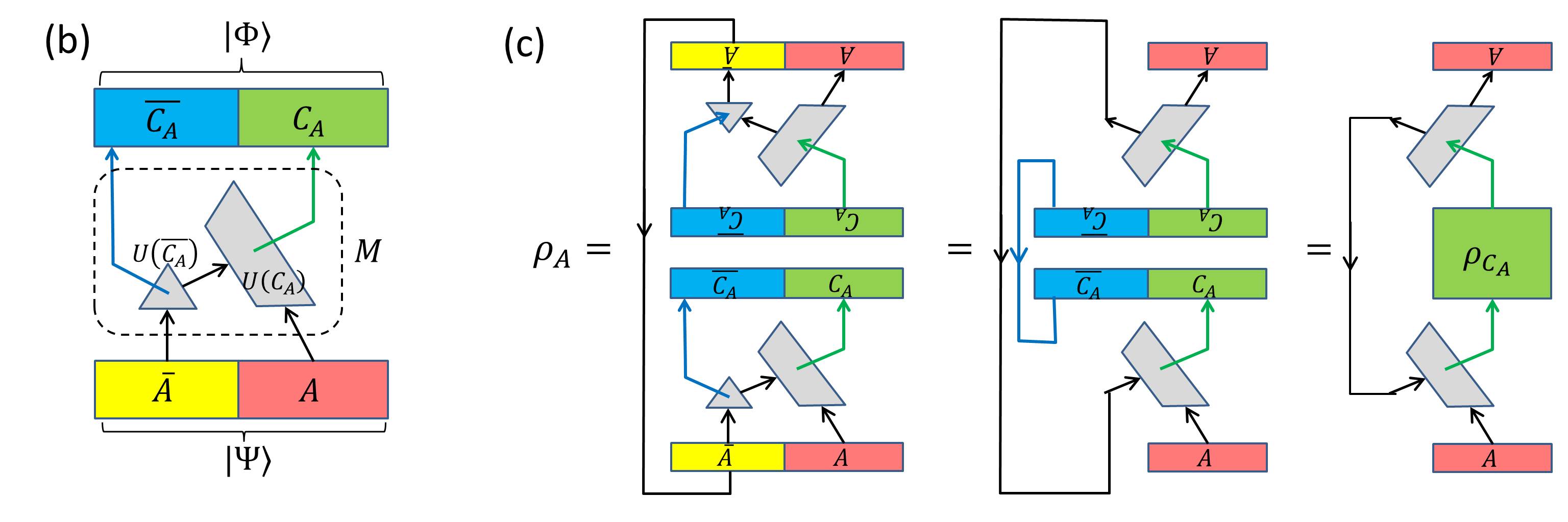}}
\caption{(a) Causal cone $C_A$ (orange region) of a boundary region $A$ (red dots). The bulk sites inside and outside the causal cone are colored green and blue, respectively. (b)  A simplified illustration of the network in (a), with regions $A,\bar{A},C_A,\overline{C_A}$ each represented by one block. The combination of all unitary transformations in the causal cone and that out of the causal cone are labeled by the tensor $U\kc{C_A}$ and $U\kc{\overline{C_A}}$, respectively. (c) Illustration of the relation between the boundary reduced density matrix $\rho_A$ and the bulk reduced density matrix $\rho_{C_A}$.} \label{fig:causalcone}
\end{figure*}

Since the EHM is an exact mapping, the causal cone structure for special MERA states does not apply. However, there is a generalized causal cone structure, as is illustrated in Fig. \ref{fig:causalcone}. Each tensor stands for the unitary transformation $U$ which maps the two incoming indices to one outgoing auxiliary index and one bulk index (blue line with a solid circle). One can draw a bulk region that has $A$ as its boundary, and has only incoming arrows acrossing it. For such a region, the inverse of EHM maps the bulk states in this region to boundary degrees of freedom in $A$ and auxiliary degrees of freedom, without replying on other degrees of freedom outside this region.  The causal cone $C_A$ is defined as the minimal one among all such bulk regions.

Consider a boundary state $\ket{\Phi}$ which is related to a bulk state $\ket{\Psi}$ by the EHM $\ket{\Phi}=M^{-1}\ket{\Psi}$. Now we want to obtain the reduced density matrix of a region $A$ on the boundary. $M$ consists of a sequence of unitary transformations. As is illustrated in Fig. \ref{fig:causalcone} (b) and (c), all the transformations outside a causal cone cancels each other in the partial trace, so that $\rho_A={\rm tr}_{\bar{A}}\ket{\Phi}\bra{\Phi}$ is determined by the reduced density matrix of the bulk state in the causal cone $C_A$:
\bea
\rho_{C_A}&=&{\rm tr}_{\overline{C_A}}\ket{\Psi}\bra{\Psi}\nn\\
\rho_{A}&=&{\rm tr}_{\rm aux.}\kd{M\kc{C_A}^{-1}\rho_{C_A}M\kc{C_A}}
\eea
Here $\bar{A}$ is the complementary set of $A$ in the boundary, and $\overline{C_A}$ is that of $C_A$ in the bulk. $M\kc{C_A}$ is the unitary transformations in the causal cone, which maps the bulk states in $C_A$ to auxiliary sites and boundary sites in $A$. The density matrix of $A$ is obtained by tracing out the auxiliary sites.

Therefore we see that the computation of the boundary reduced density matrix still only involve $\sim \log L$ number of bulk sites. In general the bulk reduced density matrix $\rho_{C_A}$ cannot be obtained, which forbids us to make use of the causal cone structure. However if we take some ansatz states in the bulk such as a free fermion state, or a tensor product state (TPS), it is possible to obtain $\rho_{C_A}$ and calculate the boundary reduced density matrix. It should be noticed that the boundary state can be interacting even if the bulk state is a free fermion state, if the mapping $V$ at each vertex of the network does not preserve the quadratic nature of the Hamiltonian. In this point of view, EHM with short-range entangled bulk states can be taken as a larger class of variational states, which generalizes MERA states and in general allows a better approximation to the boundary ground state.

\subsection{Comparison of EHM and the ordinary AdS/CFT duality}

A natural question is what is the relation of EHM with ordinary AdS/CFT duality. In particular, it has been proposed\cite{klebanov2002,sezgin2005} that a free boson or free fermion $O(N)$ vector model is dual to the Vasiliev theory\cite{vasiliev1992}, which contains interaction between infinite number of high spin fields, one at each spin. (For real bosons (Majorana fermions), only even-spin (odd-spin) fields are present. If we believe that the free fermion in continuum can be viewed as a continuum limit of the lattice Dirac fermion studied in this paper, there appears to be a contradiction since EHM leads to a free fermion theory in the bulk rather than Vasiliev theory. One possible explanation of this contradiction is that there is no continuum limit of the bulk theory we obtained by EHM. However, there seems to be no principle to exclude the analog of EHM in continuum systems. Assuming such a mapping for free fermion can be found,  it will be a unitary transformation of the fermion field
\bea
\eta_a\kc{X}=\int d^dyM\kc{X|y}\psi_a(y)\label{continuousEHM}
\eea
Here $\psi_a(y)$ is a boundary fermion field and $\eta_a\kc{X}$ is a bulk fermion field, with $a=1,2,...,N$ an $O(N)$ index. $X$ and $y$ are bulk and boundary coordinates, respectively. The space of arbitrary field configurations on AdS$_{d+1}$ is much higher dimensional than that on R$^d$, but it may be possible to define a unitary mapping between the field configurations with a suitable UV cut-off, as is indicated by the lattice EHM.

Assuming such a mapping is possible, we can obtain a quadratic fermion action $S_{\rm bulk}\kd{\eta_a,\bar{\eta}_a}$ in the bulk from the quadratic action of the boundary fermion $\psi_a(y)$. It should be noticed that the bulk theory obtained in this way contains all states of the free fermion system on the boundary, including the states that are not $O(N)$ invariant. This is the key difference from the Vasiliev theory which only contain fields that correspond to $O(N)$ invariant single-trace operators on the boundary. If we introduce the most generic $O(N)$ invariant single-trace source term for the bulk fermion, we can define
\bea
Z\kd{J\kc{X,X'}}&=&\exp\kd{-S_{\rm bulk}\kd{\eta_a,\bar{\eta}_a}\right.\nn\\
& &\left.+\int d^{d+1}Xd^{d+1}X'J\kc{X,X'}\bar{\eta}_a\kc{X'}\eta_a(X)}\nn\\
\eea
This defines the action of the bilocal field $J\kc{X,X'}$ by
\bea
S_{\rm eff}\kd{J\kc{X,X'}}=-\log Z\kd{J\kc{X,X'}}\label{Seffbilocal}
\eea
This effective action encodes all $O(N)$ invariant correlation functions of the bulk fermion (and thus the boundary fermion). Although we haven't worked out this procedure sketched above explicitly, we would like to make the conjecture that for a suitable choice of the mapping $M(X|y)$ defined in Eq. (\ref{continuousEHM}), action (\ref{Seffbilocal}) reproduces the Vasiliev theory when the bilocal field $J\kc{X,X'}$ is expanded into different spin components. Physically, this conjecture means that the strong interaction in Vasiliev theory comes from the simple fact that we insist to study the $O(N)$ singlet sector of a free fermion or free boson theory, while the well-defined propagating modes in this system are $O(N)$ vectors. This is similar to what happens when one tries to describe the particle-hole excitations of a Fermion system in space-time dimension higher than $2$. When there is no well-defined collective mode (such as spin waves), one ends up with a large number of boson fields interacting with each other.

Another theory that EHM shall be compared with is the theory of S.-S. Lee\cite{lee2010,lee2011,lee2012}, which also constructs the bulk theory by modifying an RG procedure of the boundary theory. Similar to Vasiliev theory, what is obtained for $O(N)$ vector model in Ref.\cite{lee2010} is an interacting theory that describes the $O(N)$ singlet sector.

\subsection{Some more thoughts on the space-time geometry}\label{sec:generals}

In the approach so far, we have considered a fixed tree-like background network, and define a distance on this network by two-point correlation functions. There are apparently many open questions in this scheme. Taking the spatial distance defined in Eq. (\ref{distancedef}) as an example. To make this distance $d_{\bf xy}=-\xi \log\frac{I_{\bf xy}}{I_0}$ a legitimate distance function, the triangle inequality needs to be satisfied, which means the mutual information should satisfy
\bea
I_{\bf xy}I_{\bf yz}\leq I_{\bf xz}
\eea
for any three points in the bulk. Apparently this is not always true for a generic state. Physically this equation requires some locality of correlation and entanglement, {\it i.e.}, the correlation between two farther away points ${\bf x}$ and ${\bf z}$ are mediated by the third point ${\bf y}$ which is on the shortest path between ${\bf x}$ and ${\bf z}$. The identification of correlation/entanglement with geometrical distance should be only made in the large scale (long distance) limit, but it isn't clear how to define this condition more quantitatively.

%We suspect that this equation holds when i) the bulk is a short-range entangled state, and ii) we only study the large scale geometry in the scale much longer than the correlation length $\xi$. The proof of this conjecture will be left for future work. %In a different point of view, we can give a real space path integral representation to a correlation function $C_{\bf xz}=\avg{\hat{O}_{\bf x}\hat{O}_{\bf z}}=\sum_{L\kc{\bf xz}}P_L$, with the sum over all paths $L\kc{\bf xz}$ connecting ${\bf x}$ and ${\bf z}$. If the path integral is in the semiclassical limit where one path dominates the sum, the correlation function

In general, there is no reason to view the bulk geometry as a static classical background. The distance function calculated in this work should be considered as an average distance in a certain coordinate choice, for a fluctuating bulk geometry. For the same boundary theory, it is possible that different EHM can be defined, each of which leads to a local bulk theory. The equivalence between such bulk theories can be viewed as a large symmetry group of the bulk theory, which include gauge symmetries and general covariance of the bulk as subgroups. It should be noted that the definition of ``average distance" requires to specify a coordinate for each of the fluctuating geometry, which is therefore not general covariant.\cite{maldacenaprivate} This is consistent with the fact that the ``average distance" is defined for a particular EHM. The choice of EHM acts as a gauge fixing.

So far we have been taking a tree-like background in defining EHM. The tensor network can be viewed as a discretization of hyperbolic space, but the metric defined by correlation functions is generically different from that of the hyperbolic space, unless the boundary theory is critical. One can interpret the tensor network we start with as a classical background geometry, and consider the emergent metric defined by correlation functions as a quantum correction to the geometry. (For the special states defined in MERA, the quantum correction vanishes.) Generically, the hyperbolic space we start with is not a ``saddle point" so that the correction leads to a different geometry. There is no particular reason to start from the hyperbolic space. To avoid relying on the specific starting point, one can consider EHM on a more generic network, and determine the network self-consistently. The self-consistent equation is determined by the condition
\bea
-\xi \log C_{\bf xy}=d_{\bf xy}=d_{\bf xy}^g\label{SCE}
 \eea
where $d_{\bf xy}$ is the distance defined from correlation function $C_{\bf xy}$, and $d_{\bf xy}^g$ is the graph distance on the network. (We assume that the unitary transformations on all vertices of the network are identical, so that all information about the geometry is in the network itself.) An interesting question is whether the ``fix point" space-time geometry determined by the self-consistent equation (\ref{SCE}) satisfies Einstein equation.

\section{Conclusion}

In conclusion, we have proposed an exact holographic mapping between $d$-dimensional and $d+1$-dimensional quantum many-body states. For suitable mapping the $d+1$-dimensional bulk theory is local and short-range entangled, and we can use the bulk correlation function to define the emergent bulk geometry. In this case, the bulk geometry is a ``holographic dual description" of the boundary $d$-dimensional theory. The general idea of EHM is to find a new direct-product decomposition of the Hilbert space, in which entanglement between different sites is short-ranged even if in the original system the correlation and entanglement may be long-ranged. It is such ``quasi-local" basis which defines a ``geometrized" description of the system.

 For the example of $1+1$-d free fermions, we studied the bulk geometry corresponding to several different systems, including massive and massless fermions at zero temperature, and massless fermions at finite temperature. The bulk geometry obtained is qualitatively consistent with the expectation from AdS/CFT duality. In particular, for a finite temperature system we show that the IR region of the network behaves like the near-horizon region of a black-hole. As an example of time-dependent geometry, we studied the quantum quench problem in two $1+1$-d chains. In the initial state, the two chains are entangled and the bulk geometry is a wormhole geometry with two asymptotic AdS regions. After the quench the two chains are decoupled and the wormhole shrinks and stretches. This bulk-edge correspondence is also consistent with the known results in AdS/CFT, except that the system dethermalizes in a short time proportional to the system size, so that the wormhole size will oscillate. This is an artifact of free fermion systems, due to infinite number of conserved quantities.

 A lot of open questions remain to be studied in this new scheme. We discussed that EHM has a similar causal cone structure as MERA which allows the numerical calculation of boundary properties corresponding to certain simple bulk states. A general question is how to understand the bulk geometry in a more complete and background independent way. We discussed the possibility of choosing the bulk geometry self-consistently. This can be viewed as a ``mean-field approximation" of a fluctuating geometry. Another open question is whether we should generalize the definition of ``bulk geometry" to include the information about more generic correlation functions, rather than just two-point correlation. It is also interesting to study black hole physics using this new approach. In particular, one may wonder whether it is possible to create a black hole with Hawking radiation, such that the black hole information parodox\cite{hawking1976,susskind1993,stephens1994} can be tested. These open questions will be the topics of future research.

\noindent{\bf Acknowledgement.} We acknowledge helpful discussion with Sean Hartnoll, Chaoming Jian, Juan Maldacena, Shinsei Ryu, Brian Swingle, T. Senthil, Frank Verstraete, Xiao-Gang Wen, Edward Witten, and in particular Leonard Susskind and Guifre Vidal. This work is supported by the National Science Foundation through the grant No. DMR-1151786.

\bibliography{holography}

\begin{thebibliography}{38}
\bibitem[{mal()}]{maldacenaprivate}
\bibinfo{note}{I would like to thank Juan Maldacena for private communications
  about this point.}
\end{thebibliography}

\begin{widetext}

\appendix
\section{The detail of the EHM for free fermion model}

\subsection{Bulk Green's function}

%the mapping, basis and Hamiltonian recurrence relation.
The mapping to the operators is defined by the two equations (\ref{holomapping}) and (\ref{holomapping2}). From these two equations it is easy to see that
\bea
a_{jn}&=&\frac1{\sqrt{2}}\kc{a_{2j-1,n-1}+a_{2j,n-1}}=\frac12\sum_{l=4j-3}^{4j}a_{l,n-2}=...=2^{-n/2}\sum_{l=2^n(j-1)+1}^{2^nj}c_{l}
\eea
Therefore $a_{jn}^\dagger$ creates a state with square shape wavefunction, which is a constant at the $2^n$ sites from $2^n(j-1)+1$ to $2^nj$, and zero elsewhere. From Eq. (\ref{holomapping2}) we can then obtain the bulk state $b_{jn}$:
\bea
b_{jn}&=&2^{-n/2}\kd{-\sum_{l=2^n(j-1)+1}^{2^n(j-1)+2^{n-1}}c_{l}+\sum_{l=2^n(j-1)+2^{n-1}+1}^{2^n j}c_{l}}\equiv\sum_l\phi_{in}(l)c_l
\eea
The last step is a definition of the wavefunction $\phi_{jn}(l)$, which is known as the Haar wavelet\cite{haar1910}. By a Fourier transformation we obtain
\bea
b_{jn}&=&\sum_{q=2\pi n/2^N,~n=1,2,...,2^N}\phi_{jn}^*(q)c_q\nn\\
\phi_{jn}(q)&=&2^{-N/2}\sum_l\phi_{jn}(l)e^{-iql}
\eea
The explicit form of $\phi_{jn}(q)$ can be obtained. From
\bea
\phi_{jn}(l+1)-\phi_{jn}(l)&=&2^{-n/2}\kd{-\delta_{l,2^n(j-1)}-\delta_{l,2^nj}+2\delta_{l,2^nj-2^{n-1}}}
\eea
we obtain
\bea
\phi_{jn}(q)\kc{e^{iq}-1}&=&2^{-N/2}2^{-n/2}\kd{-e^{-iq2^n(j-1)}-e^{-iq2^nj}+2e^{-iq\kc{2^nj-2^{n-1}}}}\nn\\
\=>\phi_{jn}(q)&=&-\frac{2^{-(n+N)/2}e^{-iq2^nj}\kc{e^{iq2^{n-1}}-1}^2}{e^{iq}-1}
\eea
The bulk Green's function is thus
\bea
\avg{Tb_{jn\alpha}(\tau) b_{km\beta}^{\dagger}}&=&\sum_q\phi_{jn}^*(q)\phi_{km}(q)G_{q\alpha\beta}\\
\text{with~}G_{\alpha\beta}(q)&=&\avg{Tc_{q\alpha}(\tau)c_{q\beta}^\dagger(0)}\nn
\eea
The boundary Green's function can be explicitly written (in $2\times 2$ matrix form) as
\bea
G_q&=&e^{-\tau h_q}\kc{1+e^{-\beta h_q}}^{-1}
\eea
for $\tau\in(0,\beta]$.  For the Dirac Hamiltonian
\bea
h_q&=&\left[ \sigma_x\sin q+\kc{m+B\kc{1-\cos q}}\sigma_y\right]
\eea
the expression can be further simplified to
\bea
G_q&=&\frac12\kc{\cosh\kc{\tau E_q}\eye-\frac{h_q}{E_q}\sinh \kc{\tau E_q}}\kc{\eye+\frac{h_q}{E_q}\tanh\frac{\beta E_q}2}\label{GreenDirac}
\eea

\subsection{``Renormalization" of the Hamiltonian}

For a generic Hamiltonian rather than the lattice Dirac model (\ref{HDirac}), one can still apply the same EHM to obtain a bulk theory. Since the auxiliary fermions $a_{j,n+1}$ is obtained from transformation of $a_{jn}$, the low energy quadratic Hamiltonian in the $n+1$-th layer $h_a^{(n+1)}$ is completely determined by $h_a^{(n)}$. Therefore we can write a generic iterative relation between $h_a^{(n+1)}$ and $h_a^{(n)}$, which plays the role of RG equation. For simplicity, we consider translation invariant Hamiltonians. We start from the $a$ Hamiltonian in $n$-th layer
\bea
H_a^{(n)}=\sum_k a_{k,n}^\dagger h_{ak}^{(n)}a_{k,n}
\eea
The transformation \ref{holomapping2} can be Fourier transformed to
\bea
a_{q,n+1}&=&2^{-\kc{N-n-1}/2}\sum_{j=1}^{2^{N-n-1}}a_{j,n+1}e^{-iqj}
=2^{-\kc{N-n}/2}\sum_{j=1}^{2^{N-n-1}}\kc{a_{2j-1,n}+a_{2j,n}}e^{-iqj}\nn\\
&=&2^{-\kc{N-n}}\sum_{j=1}^{2^{N-n-1}}\sum_{p}a_{p}e^{i(2p-q)j}\kc{e^{-ip}+1}\nn\\
&=&\frac{1+e^{-iq/2}}2a_{q/2}+\frac{1-e^{-iq/2}}2a_{q/2+\pi}
\eea
Similarly
\bea
b_{q,n+1}&=&2^{-\kc{N-n}}\sum_{j=1}^{2^{N-n-1}}\sum_{p}a_{p}e^{i(2p-q)j}\kc{-e^{-ip}+1}\nn\\
&=&\frac{1-e^{-iq/2}}2a_{q/2}+\frac{1+e^{-iq/2}}2a_{q/2+\pi}
\eea
The Hamiltonian can be rewritten as
\bea
H_a^{(n)}&=&\sum_{k\in[0,\pi)}\kc{a_{kn}^\dagger~a_{k+\pi,n}^\dagger}\kc{\ba{cc}h_{ak}^{(n)}&\\
&h_{a,k+\pi}^{(n)}\ea}\vect{a_{kn}\\a_{k+\pi,n}}\nn\\
&=&\sum_{q\in[0,2\pi)}\kc{a_{q,n+1}^\dagger~b_{q,n+1}^\dagger}V^\dagger\kc{\ba{cc}h_{aq}^{(n)}&\\
&h_{a,q/2+\pi}^{(n)}\ea}V\vect{a_{q,n+1}\\b_{q,n+1}}\\
\text{with~}V&=&\kc{\ba{cc}\frac{1+e^{iq/2}}2&\frac{1-e^{iq/2}}2\\\frac{1-e^{iq/2}}2&\frac{1+e^{iq/2}}2\ea}\nn
\eea
Therefore $H_a^{(n+1)}$ is determined by the upper block of the transformed Hamiltonian:
\bea
h_{aq}^{(n+1)}=h_{a,q/2}^{(n)}\frac{1+\cos\frac{q}2}2+h_{a,q/2+\pi}^{(n)}\frac{1-\cos\frac{q}2}2\label{RGE}
\eea
Eq. (\ref{RGE}) plays the role of RG equation of the Hamiltonian. Since the transformation does not act on spin, each component of the Hamiltonian satisfies this equation. It can be checked that for a Hamiltonian of the form $h_{a,q}^{(n)}=\sin qA_1+(1-\cos q)A_2$, with $A_1,A_2$ arbitrary matrices independent of $q$, we obtain $h_{aq}^{(n+1)}=\frac12h_{aq}^{(n)}$. This result shows that the lattice mapping is different from an RG flow in continuum limit, since in the latter case the scaling dimension of the term $1-\cos k\simeq k^2/2$ will be different from that of $\sin k\simeq k$.

\section{The detail of the fitting procedure}
\subsection{The fitting of zero temperature results}

The coordinates of the sites are given by Eq. (\ref{horizontald}) for the critical system, which is determined by scaling and translation symmetries. The geodesic distance in (Euclidean) AdS space can be written simply in the embedded coordinates $X^a,~a=1,2,3,4$. In this coordinate the AdS space is a hyperbolic surface embedded in the 4d flat space with Lorentz metric, determined by the equation $X^aX^b\eta_{ab}=R^2$. Here $\eta_{ab}={\rm diag}[1,-1,-1,-1]$ is the Lorentz metric. The relation between $X^a$ and the intrinsic coordinate $\rho,\theta,t$ is
\bea
X=\kc{\sqrt{\rho^2+R^2}\cosh\frac {t}{R},\sqrt{\rho^2+R^2}\sinh\frac {t}R,\rho\cos\theta,\rho\sin\theta}\label{embeddedAdS}
\eea
The geodesic distance is
\bea
d_{X_1X_2}=R{\rm acosh}\kc{\frac{X_1^aX_2^b\eta_{ab}}{R^2}}\label{geodesicAdSembedded}
\eea

For two sites $(x,n),(y,n)$ separated horizontally, the distance reduces to formula (\ref{horizontald}). We have
\bea
\log \frac{I_0}{I_{(x,n),(y,n)}}=\frac{d_{(x,n),(y,n)}}{\xi}\simeq \frac{2R}{\xi}\log \frac{|x-y|}{R},~\text{for~}\abs{x-y}\gg R
\eea
Since $\xi$ is unknown, we first do a linear fitting at large $|x-y|$ to obtain
\bea
\frac{d_{(x,n),(y,n)}}{\xi}\simeq P_0+P_1\log\abs{x-y}
\eea
Then we obtain $R$ from $\frac{P_1}{P_0}=-\log R$ and
and then obtain $\xi$ from $P_0=2R/\xi$.

Now we input the $R$ value to Eq. (\ref{horizontald}) to obtain the AdS distance between points $(x,1)$ and $(x,n)$ separated in radial direction. By a linear fitting of this distance with the $\frac{d}{\xi_\perp}=\log\frac{I_0}{I_{(x,1),(x,n)}}$ we can obtain the correlation length $\xi_\perp$. As is shown in Fig. \ref{MIcritical} (b), $\xi_{\perp}$ is different from $\xi$ in the horizontal direction.

The distance between two points $(\rho,\theta,0)$ and $(\rho,\theta,\tau)$ is
\bea
d(\rho,\tau)=R~{\rm acosh}\kd{\kd{\frac{\rho^2}{R^2}+1}\cosh\frac {t}{R}-\frac{\rho^2}{R^2}}
\eea
However, there is a rescaling of time $t$ that we need to include in comparison with the boundary system.
At the boundary $\rho=\frac{L}{2\pi}$ ($L=2^N$ is the perimeter), and the metric reduces to
\bea
ds^2=\kc{\frac{L^2}{4\pi^2R^2}+1}dt^2+\rho^2d\theta^2
\eea
Therefore we should rescale $t\>t/ \sqrt{\frac{L^2}{4\pi^2l^2}+1}$ so that at the boundary we have the standard metric $dt^2+\rho^2d\theta^2$ (with speed of light $c=1$). After the rescaling the geodesic distance is
\bea
d(r,t)=R{\rm acosh}\kd{\kd{\frac{\rho^2}{R^2}+1}\cosh\frac {t}{\sqrt{\frac{L^2}{4\pi^2}+R^2}}-\frac{\rho^2}{ R^2}}
\eea
Consider the limit
\bea
\rho\gg 2\pi R,~\frac{L}{2\pi}\gg t\gg R
\eea
which leads to
\bea
d(r,t)\simeq 2R\log\frac{2\pi\rho t}{RL}
\eea
Using this formula, the same fitting procedure as the spatial distance leads to an independent way to determine $R$ and $\xi$. However, it should be noted that the time-direction distance is defined by the single-particle Green's function, so one does not expect the $\xi$ to be compared with that observed in spatial distance.

\subsection{The fitting of finite temperature results}

A special property of gravity in 3-dimension is that there is a black-hole solution, the BTZ solution, which is a quotient of the AdS space. In other words, the black-hole solution is locally equivalent to AdS$_3$.
The quotient can be seen in the following parameterization of the embedded coordinate
\bea
X&=&R\kc{\frac \rho b\cosh\kc{ \frac{b}R\theta},\frac \rho b\sinh \kc{\frac{b}R\theta},
\sqrt{\frac{\rho^2}{b^2}-1}\sin \kc{\frac{bt}{R^2}},
\sqrt{\frac{\rho^2}{b^2}-1}\cos \kc{\frac{bt}{R^2}}}
\eea
Compare this expression with the pure AdS$_3$ case (\ref{embeddedAdS}) we see that the black-hole solution is obtained by a double Wick rotation from the pure AdS$_3$ solution $t\>b\theta, \theta\>\frac{bt}{R^2}$ and replace $\rho\>R\sqrt{\frac{\rho^2}{b^2}-1}$. After the rotation, time $t$ is periodic with periodicity $\beta=\frac{2\pi R^2}{b}$, and $\theta$ becomes a real number. We then compactify the $\theta$ direction by identifying the points $\theta$ with $\theta+2n\pi,~n\in\mathbb{Z}$, which can also be viewed as taking the quotient of AdS space to a $\mathbb{Z}$ subgroup of the isometry group $SO(2,1)$. The metric in the intrinsic coordinates $\rho,\theta,t$ is
\bea
ds^2=\frac{\rho^2-b^2}{R^2}dt^2+\frac{R^2}{\rho^2-b^2}d\rho^2+\rho^2d\theta^2
\eea
$b$ has the physical meaning of black-hole radius, which also determines the temperature. It should be noticed that the time needs to be rescaled when compared with the boundary system, in the same way as in the zero temperature case. The rescaling is defined as
\bea
t\>t\frac{R}{\sqrt{\rho_0^2-b^2}}=t\frac{R}{\sqrt{\frac{L^2}{4\pi^2}-b^2}}
\eea
After the rescaling, the period of the boundary time is
\bea
\beta=2\pi R\sqrt{\frac{L^2}{4\pi^2b^2}-1}\label{BHtemperature}
\eea
which is the inverse temperature of the boundary system.

Now we look at the time-direction distance between two points $(\rho,\theta,t=0)$ and $(\rho,\theta,t)$. The distance can still be computed by the AdS formula (\ref{geodesicAdSembedded})
\bea
d_{(\rho,\theta,0),(\rho,\theta,t)}&=&R~{\rm acosh}\kd{\frac{\rho^2}{b^2}-\kc{\frac{\rho^2}{b^2}-1}\cos\kc{\frac{2\pi}\beta t}}\label{dfiniteT}
\eea
Taking $t=\beta/2$, we obtain the ``maximal" in the time circle as
\bea
d_{\rm max}(\rho)=R{\rm acosh}\kd{\frac{2\rho^2}{b^2}-1}=2R{\rm acosh}\kc{\frac{\rho}{b}}\label{dmax}
\eea
The numerically obtained time-direction correlation function (\ref{TimeCor}) shall be fitted with the analytic formula
\bea
-\log\frac{C_{\bf x}\kc{t}}{C_0}=\frac{d_{(\rho,\theta,0),(\rho,\theta,t)}}{\xi}\kc{\frac{R}{\xi},\frac{\rho}{b}}
\eea
The righthand side means that $\frac{d_{(\rho,\theta,0),(\rho,\theta,t)}}\xi$ is a function of two dimensionless parameters $\frac{R}{\xi}$ and $\frac{\rho}{b}$. The constant $C_0=C_{\bf x}(t=0)$ is the trace of the equal time correlation function, which is $1$ as can be seen from Eq. (\ref{GreenDirac}). To determine the parameters $b,R,\xi$, we first take $\frac{R}{\xi}$ as a parameter and obtain $\frac{\rho}b$ as a function of $\frac{R}\xi$ from the maximal distance in Eq. (\ref{dmax}). By inputting this $\frac{\rho}{b}$ value to Eq. (\ref{dfiniteT}), we obtain the distance $d_{(\rho,\theta,0),(\rho,\theta,t)}=d\kc{\frac{R}{\xi}}$ as a function of the single parameter $\frac{R}{\xi}$.  Then we compare the resulting distance function and determine the optimal $\frac{R}{\xi}$ by minimizing the square-averaged deviation function
\bea
\delta^2=\sum_{\bf x}\kd{\log\frac{C_{\bf x}\kc{t}}{C_0}+\frac{d_{(\rho,\theta,0),(\rho,\theta,t)}}{\xi}\kc{\frac{R}{\xi}}}^2
\eea
To determine the value of $b,R,\xi$ we use the boundary condition of $\rho$. By an linear extrapolation of the function $\log\kc{\rho_n/b}$ as a function of $n$, we obtain $\rho_0/b$ at $n=0$. $\rho_0$ is the boundary value of $\rho$ which should be identified with $\rho_0=L/2\pi$. This determines the value of $b$. Then we determine $R$ by $b$ and temperature $T$ from Eq. (\ref{BHtemperature}). Once $R$ is determined, $\xi=R/(R/\xi)$ can be obtained.

\end{widetext}
\end{document}